\documentclass[preprint,aps,12pt,showpacs,nofootinbib,tightenlines]{revtex4}
\usepackage{amsmath}
\usepackage{amssymb}
\usepackage{epsfig}
\usepackage{graphicx}
\begin{document}

\newcommand{\beq}{\begin{eqnarray}}
\newcommand{\eeq}{\end{eqnarray}}
\newcommand{\non}{\nonumber\\ }

\newcommand{\acp}{{\cal A}_{CP}}
\newcommand{\calm}{{\cal M} }
\newcommand{\cala}{{\cal A} }
\newcommand{\calh}{{\cal H} }
\newcommand{\ov  }{\overline  }

\newcommand{\etap}{\eta^{(\prime)} }
\newcommand{\pb}{\phi_B}

\newcommand{\pka}{\phi_{K}^A}
\newcommand{\pepa}{\phi_{\eta'}^A}
\newcommand{\pkp}{\phi_K^P}
\newcommand{\pep}{\phi_{\eta}^P}
\newcommand{\pkpp}{\phi_{\eta'}^P}
\newcommand{\pkt}{\phi_K^T}
\newcommand{\pet}{\phi_{\eta}^T}
\newcommand{\pkpt}{\phi_{\eta'}^T}
\newcommand{\pksa}{\phi_{K^*}}
\newcommand{\pksp}{\phi_{K^*}^s}
\newcommand{\pkst}{\phi_{K^*}^t}
\newcommand{\fb}{f_B }
\newcommand{\fk}{f_K }
\newcommand{\fe}{f_{\eta} }
\newcommand{\fep}{f_{\eta'} }
\newcommand{\rks}{r_{K^*} }
\newcommand{\rk}{r_K }

\newcommand{\mb}{m_B }
\newcommand{\mw}{m_W }

\newcommand{\xeba}{\bar{x}_2}
\newcommand{\xsba}{\bar{x}_3}
\newcommand{\res}{r_{\eta_{s\bar{s}}}}
\newcommand{\red}{r_{\eta_{d\bar{d}}}}
\newcommand{\pkas}{\phi^A_{\eta_{s\bar{s}}}}
\newcommand{\pkps}{\phi^P_{\eta_{s\bar{s}}}}
\newcommand{\pets}{\phi^T_{\eta_{s\bar{s}}}}
\newcommand{\pkad}{\phi^A_{\eta_{d\bar{d}}}}
\newcommand{\pkpd}{\phi^P_{\eta_{d\bar{d}}}}
\newcommand{\petd}{\phi^T_{\eta_{d\bar{d}}}}
\newcommand{\pvsl}{ p \hspace{-2.0truemm}/_{K^*} }
\newcommand{\esl}{ \epsilon \hspace{-2.1truemm}/ }
\newcommand{\psl}{ p \hspace{-2.0truemm}/ }
\newcommand{\nsl}{ n \hspace{-2.2truemm}/ }
\newcommand{\vsl}{ v \hspace{-2.2truemm}/ }
\newcommand{\epsl}{\epsilon \hspace{-1.8truemm}/\,  }
\newcommand{\bfkk}{{\bf k} }

\def \epjc{ Eur. Phys. J. C }
\def \jpg{  J. Phys. G }
\def \npb{  Nucl. Phys. B }
\def \plb{  Phys. Lett. B }
\def \pr{  Phys. Rep. }
\def \prd{  Phys. Rev. D }
\def \prl{  Phys. Rev. Lett.  }
\def \zpc{  Z. Phys. C  }
\def \jhep{ J. High Energy Phys.  }
\def \ptp{ Prog. Theor. Phys.  }

\title{NLO contributions to $B \to K K^*$ Decays in the pQCD approach }
\author{Zhi-Qing Zhang \footnote{Electronic address: zhangzhiqing@zzu.edu.cn}
and Zhen-Jun Xiao
\footnote{Electronic address: xiaozhenjun@njnu.edu.cn}}
\affiliation{{\it  Department of Physics and Institute of
Theoretical Physics, Nanjing Normal University, Nanjing, Jiangsu
210097, P.R.China}  \footnote{Mailing address} } 
\date{\today}
\begin{abstract}
We calculate the important next-to-leading-order (NLO) contributions
to the  $B \to K K^*$ decays from the vertex corrections, the quark
loops, and the magnetic penguins in the perturbative QCD (pQCD)
factorization approach. The pQCD predictions for the CP-averaged
branching ratios are
$Br(\ B^+ \to K^+ \overline{K}^{*0}) \approx 3.2\times 10^{-7}$,
$Br(\ B^+ \to \overline{K}^0 {K}^{*+}) \approx 2.1\times 10^{-7}$,
$Br(B^0/\ov{B}^0 \to  K^0\overline{K}^{*0}+\overline{K}^0 K^{*0})
\approx 8.5\times 10^{-7}$,
$Br(\ B^0/\ov{B}^0 \to K^+K^{*-} + K^-K^{*+}) \approx 1.3\times 10^{-7}$,
which agree well with both the experimental upper limits and the
predictions based on the QCD factorization approach. Furthermore,
the CP-violating asymmetries of the considered decay
modes are also evaluated. The NLO pQCD predictions for
$\acp(B^+ \to K^+\overline{K}^{*0})$ and
$\acp(B^+ \to K^{*+}\overline{K}^{0})$ are
 $\acp^{dir}(K^+\overline{K}^{*0})\approx -6.9 \%$ and
$\acp^{dir}(K^{*+}\overline{K}^0)\approx 6.5 \%$.
\end{abstract}

\vspace{1cm}


\maketitle

\section{Introduction}

It is well-known that the experimental measurements and theoretical studies
of the two body charmless hadronic B meson decays play an important role in the
precision test of the standard model (SM) and in searching for the
new physics beyond the SM \cite{cpv}.
For these decays, the dominant theoretical error comes from the large
uncertainty in evaluating the so-called hadronic matrix element,
$\langle M_1 M_2|O_i|B\rangle$, where $M_1$ and $M_2$ are light final state
mesons.  The perturbative QCD (pQCD) approach \cite{li2003}
is one of the most popular factorization approaches \cite{bbns99,scet1} being used to
calculate the hadronic matrix elements.

When compared with the QCDF or SCET factorization approaches,
the pQCD approach has the following three special features:
(a) since the $k_T$ factorization is employed here,
the resultant Sudakov factor as well as the threshold resummation
can enable us to regulate the end-point singularities effectively;
(b) the form factors for $B \to M$ transition can be calculated perturbatively, although
some controversies still exist about this point;
and (c) the annihilation diagrams are calculable and play an important role in producing
CP violation.

Up to now, almost all two-body charmless $B/B_s \to M_1 M_2$ decays have been
calculated by using the pQCD approach at the leading order
\cite{kls01,luy01,li01,li05,liu06,guo07,bs07,bs07b}.
Very recently, some next-to-leading (NLO)
contributions to $B \to K\pi$ and several $B \to PV$ decay modes \cite{nlo05,nlopv}
have been calculated, where the
Wilson coefficients at NLO accuracy are used, and the contributions from
the vertex corrections, the quark loops and the chromo-magnetic penguin operator $O_{8g}$
have been taken into account. As generally expected, the inclusion of NLO contributions
should improve the reliability of the pQCD predictions.

In a previous paper \cite{guo07}, the authors calculated the branching ratios and CP violating
asymmetries of the $B^0/\ov{B}^0 \to K^0 \ov{K}^{*0}, \ov{K}^0 K^{*0}, K^+ K^{*-}, K^- K^{*+}$,
and $B^+ \to K^+ \ov{K}^{*0}$ and $\ov{K}^0 K^{*+}$ decays by employing the pQCD
approach at the leading order.
Following the procedure of Ref.~\cite{nlo05}, we here would like to calculate the
NLO contributions to the  $B \to K^*K$ decays by employing the low energy
effective Hamiltonian and the pQCD approach.

The remainder of the paper is organized as follows. In Sec.II, we
give a brief discussion about pQCD factorization approach. In
Sec.~III, we calculate analytically the relevant Feynman diagrams and
present the various decay amplitudes for the studied decay modes in
leading-order. In Sec.~IV, the NLO contributions from the vertex corrections, the
quark loops and the chromo-magnetic penguin amplitudes are evaluated. We show the
numerical results for the branching ratios and  CP asymmetries of $B
\to K^*K$ decays in Sec.~V. The summary and some discussions are
included in the final section.


\section{ Theoretical framework}\label{sec:f-work}

The pQCD factorization approach has been developed and
applied in the non-leptonic $B$ meson decays \cite{li2003} for some time.
In this approach, the decay amplitude is separated into soft, hard, and
harder dynamics characterized by different energy scales $(t, m_b,
M_W)$. It is conceptually written as the convolution,
\beq
{\cal A}(B \to M_1 M_2)\sim \int\!\! d^4k_1 d^4k_2 d^4k_3\ \mathrm{Tr}
\left [ C(t) \Phi_B(k_1) \Phi_{M_1}(k_2) \Phi_{M_2}(k_3)
H(k_1,k_2,k_3, t) \right ], \label{eq:con1}
\eeq
where $k_i$'s are
momenta of light quarks included in each meson, and $\mathrm{Tr}$
denotes the trace over Dirac and color indices. $C(t)$ is the Wilson
coefficient, which includes the harder dynamics at larger scale than $M_B$
scale and describes the evolution of local $4$-Fermi operators from $m_W$
down to $t\sim\mathcal{O}(\sqrt{\ov{\Lambda} M_B})$ scale,
where $\ov{\Lambda}\equiv M_B -m_b$. The function
$H(k_1,k_2,k_3,t)$ describes the four quark operator and the
spectator quark connected by
 a hard gluon whose $q^2$ is in the order
of $\ov{\Lambda} M_B$, and includes the
$\mathcal{O}(\sqrt{\ov{\Lambda} M_B})$ hard dynamics. Therefore,
this hard part $H$ can be perturbatively calculated. The function
$\Phi_{M_i}$ is the wave function which describes hadronization of the
quark and anti-quark into the meson $M_i$. While the function $H$
depends on the processes considered, the wave function $\Phi_{M_i}$ is
independent of the specific processes.

In the $B$ meson rest-frame, it is convenient to use light-cone coordinate $(p^+,
p^-, {\bf p}_{\rm T})$ to describe the meson's momenta,
\beq
p^\pm = \frac{1}{\sqrt{2}} (p^0 \pm p^3), \quad {\rm and} \quad
{\bf p}_{\rm T} = (p^1, p^2).
\eeq
Using these coordinates the $B$ meson and the two
final state meson momenta can be written as
\beq
P_B = \frac{M_B}{\sqrt{2}} (1,1,{\bf 0}_{\rm T}), \quad
P_{K^*} = \frac{M_B}{\sqrt{2}}(1,r^2_{K^*},{\bf 0}_{\rm T}), \quad
P_{K} = \frac{M_B}{\sqrt{2}} (0,1-r^2_{K^*},{\bf 0}_{\rm T}),
\eeq
respectively, here $r_{K^*}=m_{K^*}/M_B$. The light meson ($K$) mass has been
neglected. For the $B \to K^*K$ decays considered here, only the
vector meson's longitudinal part contributes to the decays, and its
polarization vector is $\epsilon_L=\frac{M_B}{\sqrt
2M_{K^*}}(1,-r^2_{K^*},0_{\rm T})$.
Putting the anti-quark momenta in $B$, $K^*$ and $K$ mesons as $k_1$, $k_2$, and $k_3$,
respectively, we can choose
\beq
k_1 = (x_1 P_1^+,0,{\bf k}_{\rm 1T}), \quad
k_2 = (x_2 P_2^+,0,{\bf k}_{\rm 2T}), \quad
k_3 = (0, x_3 P_3^-,{\bf k}_{\rm 3T}).
\eeq
Then, the integration over $k_1^-$, $k_2^-$, and $k_3^+$ in
eq.(\ref{eq:con1}) will lead to
\beq
{\cal A}(B \to K K^*) &\sim
&\int\!\! d x_1 d x_2 d x_3 b_1 d b_1 b_2 d b_2 b_3 d b_3 \non &&
\cdot \mathrm{Tr} \left [ C(t) \Phi_B(x_1,b_1) \Phi_{K^*}(x_2,b_2)
\Phi_{K}(x_3, b_3) H(x_i, b_i, t) S_t(x_i)\, e^{-S(t)} \right ],
\quad \label{eq:a2}
\eeq
where $b_i$ is the conjugate space
coordinate of $k_{iT}$, and $t$ is the largest energy scale in
function $H(x_i,b_i,t)$. The large logarithms ($\ln m_W/t$) coming
from QCD radiative corrections to four quark operators are included
in the Wilson coefficients $C(t)$. The large double logarithms
($\ln^2 x_i$) on the longitudinal direction are summed by the
threshold resummation \cite{li02}, and they lead to $S_t(x_i)$ which
smears the end-point singularities on $x_i$. The last term,
$e^{-S(t)}$, is the Sudakov form factor which suppresses the soft
dynamics effectively \cite{li2003}. Thus it makes the perturbative
calculation of the hard part $H$ applicable at intermediate scale,
i.e., $M_B$ scale.

\subsection{ Wilson Coefficients}\label{ssec:w-c}

For $B \to K K^*$ decays, the related weak effective Hamiltonian
$H_{eff}$  with $b \to s$ transition can be written as \cite{buras96}
\beq
\label{eq:heff}
{\cal H}_{eff} = \frac{G_{F}} {\sqrt{2}} \left \{ \sum_{q=u,c}V_{qb}
V_{qs}^* \left[ C_1(\mu) O_1^q(\mu) + C_2(\mu) O_2^q(\mu)
\right] - V_{tb} V_{td}^*  \sum_{i=3}^{10} C_{i}(\mu) \,O_i(\mu) \right \} \; .
\eeq
with $G_{F}=1.166 39\times 10^{-5} GeV^{-2}$ is the Fermi constant,
and $V_{ij}$ is the CKM matrix element, $C_i(\mu)$ are the Wilson coefficients evaluated
at the renormalization scale $\mu$ and $O_i$ are the four-fermion operators. For the case of
$b \to d $ transition, simply make a replacement of $s$ by $d$ in Eq.~(\ref{eq:heff}) and in the
expressions of $O_i$ operators, which can be found easily for example in
Refs.\cite{guo07,buras96}.

In PQCD approach, the energy scale $" t"$ is chosen at the maximum value of
various subprocess scales  to suppress the higher order corrections,
which may be larger or smaller than the $m_b$ scale.
In the range of $ t < m_b $ or $t \geq m_b$, the number of active quarks is $N_f=4$ or
$N_f=5$, respectively.
For the Wilson coefficients $C_i(\mu)$ and their renormalization group (RG) running,
they are known at NLO level currently \cite{buras96}.
The explicit expressions of the LO and NLO $C_i(\mw)$ can be found easily, for example, in
Refs.~\cite{buras96,luy01}.

When the pQCD approach at leading-order are employed, the leading order Wilson
coefficients $C_i(m_W)$, the leading order RG evolution matrix $U(t,m)^{(0)}$ from
the high scale $m$ down to $t < m$ ( for details see Eq.~(3.94) in Ref.~\cite{buras96}),
and the leading order $\alpha_s(t)$ are used:
\beq
\alpha_s(t)=\frac{4\pi}{ \beta_0 \ln \left [ t^2/ \Lambda_{QCD}^2\right]},
\eeq
where $\beta_0 = (33- 2 N_f)/3$, $\Lambda_{QCD}^{(5)}=0.225 GeV$ and
$\Lambda_{QCD}^{(4)}=0.287$ GeV.

When the NLO contributions are taken into account, however,
the NLO Wilson coefficients $C_i(m_W)$, the NLO RG evolution matrix $U(t,m,\alpha)$
(for details see Eq.~(7.22) in Ref.~\cite{buras96}) and the $\alpha_s(t)$ at two-loop level
are used:
\beq
\alpha_s(t)=\frac{4\pi}{ \beta_0 \ln \left [ t^2/ \Lambda_{QCD}^2\right]}
\cdot \left \{ 1- \frac{\beta_1}{\beta_0^2 } \cdot
\frac{ \ln\left [ \ln\left [ t^2/\Lambda_{QCD}^2  \right]\right]}{
\ln\left [ t^2/\Lambda_{QCD}^2\right]} \right \},
\label{eq:asnlo}
\eeq
where $\beta_0 = (33- 2 N_f)/3$, $\beta_1 = (306-38 N_f)/3$, $\Lambda_{QCD}^{(5)}=0.225$ GeV and
$\Lambda_{QCD}^{(4)}=0.326$ GeV.

By using the input parameters as given in the Appendix, it is easy to find the numerical
values of the LO and NLO Wilson coefficients $C_i(m_b)$ for $m_b=4.8$ GeV, as listed in
Table \ref{table-1}.

\begin{table}[thb]
\begin{center}
\caption{ The numerical values of the LO and NLO Wilson coefficients $C_i(m_b)$,
$C_{7\gamma}(m_b)$ and $C_{8g}(m_b)$. }
\label{table-1}
\vspace{0.2cm}
\begin{tabular} {l|l|l|l|l|l|l}  \hline
$C_i(m_b)$ & $C_1$ & $C_2$ &$C_3$ &$C_4$ &$C_5$ & $C_6$  \\ \hline
LO         & $-0.2812$    & $1.1246$  &$0.0130$  &$-0.0278$  &$0.0080$  &$-0.0343$ \\ \hline
NLO        & $-0.1747$    & $1.0774$  &$0.0125$  &$-0.0330$  &$0.0094$  &$-0.0393$ \\ \hline \hline
$C_i(m_b)$ & $C_7/\alpha$ & $C_8/\alpha$ &$C_9/\alpha$ &$C_{10}/\alpha$ &$C_{7\gamma}$ & $C_{8g}$  \\ \hline
LO         & $ 0.1338$    & $0.0514$  &$-1.1459$  &$0.2865$  &$-0.3109$  &$-0.1481$ \\ \hline
NLO        & $-0.0032$    & $0.0305$  &$-1.2760$  &$0.2553$  &$-0.3016$  &$--$
\\ \hline
\end{tabular}\end{center}
\end{table}

\subsection{Wave Functions}\label{ssec:w-f}

The B meson is treated as a heavy-light system.
For the B meson wave function, since the contribution of $\ov{\phi}_B$ is
numerically small ~\cite{luyang}, we here only consider the
contribution of Lorentz structure
\beq
\Phi_B= \frac{1}{\sqrt{2N_c}}
(\psl_B +m_B) \gamma_5 \phi_B ({\bf k_1}), \label{bmeson}
\eeq
with
\beq
\phi_B(x,b)&=& N_B x^2(1-x)^2 \mathrm{exp} \left
 [ -\frac{M_B^2\ x^2}{2 \omega_{b}^2} -\frac{1}{2} (\omega_{b} b)^2\right],
 \label{phib}
\eeq
where $\omega_{b}$ is a free parameter and we take
$\omega_{b}=0.4\pm 0.04$ GeV in numerical calculations, and
$N_B=101.445$ is the normalization factor for $\omega_{b}=0.4$.

The K and $K^*$  mesons are all treated as a light-light system.
The wave function of $K$ meson is defined as \cite{ball98}
\beq
\Phi_{K}(P,x,\zeta)\equiv \frac{1}{\sqrt{2N_C}}\gamma_5 \left [ \psl
\phi_{K}^{A}(x)+m_0^{K} \phi_{K}^{P}(x)+\zeta m_0^{K} (\vsl \nsl -
v\cdot n)\phi_{K}^{T}(x)\right ],
\eeq
where $P$ and $x$ are the
momentum and the momentum fraction of $K$, respectively. The
parameter $\zeta$ is either $+1$ or $-1$ depending on the assignment
of the momentum fraction $x$. For the considered
$B \to K K^*$ decays, $K^*$ meson is longitudinally polarized, and only the
longitudinal component  $\phi_{K^*}^L$ of the wave function contribute
\cite{ball98}
\beq
\phi_{K^*}^L=\frac{1}{\sqrt{2N_c}}\left\{\esl_L
\left[m_{K^*} \phi_{K^*}(x)+ \pvsl\phi_{K^*}^t(x)\right]+m_{K^*} \phi^s_{K^*}(x)\right\},
\eeq
where the first term is the leading twist wave function (twist-2),
while the second and third term are sub-leading twist (twist-3) wave
functions.

The expressions of the relevant distributions functions are the following
\cite{ball98}:
\begin{eqnarray}
 \pka(x) &=&  \frac{f_K}{2\sqrt{2N_c} }  6x (1-x)
    \left[1+a_1^{K}C^{3/2}_1(t)+a^{K}_2C^{3/2}_2(t)+a^{K}_4C^{3/2}_4(t)
  \right],\label{piw1}\\
 \pkp(x) &=&   \frac{f_K}{2\sqrt{2N_c} }
   \left[ 1+(30\eta_3-\frac{5}{2}\rho^2_{K})C^{1/2}_2(t)
   -3\left[ \eta_3\omega_3+\frac{9}{20}\rho^2_K
   (1+6a^K_2)\right]C^{1/2}_4(t)\right], \ \ \\
 \pkt(x) &=&  - \frac{f_K}{2\sqrt{2N_c} } t
   \left[ 1+6(5\eta_3-\frac{1}{2}\eta_3\omega_3-\frac{7}{20}\rho^2_{K}
   -\frac{3}{5}\rho^2_Ka_2^{K})
   (1-10x+10x^2)\right] ,\quad\quad\label{piw}
 \end{eqnarray}
with the mass ratio $\rho_K=m_K/m_{0K}$, and $\eta_3=0.015$, $\omega=-3.0$.
Since the uncertainties of the currently available Gegenbauer moments \cite{ball05}
are still large, we vary the value of $a_K^1$ and $a_K^4$ by $100\%$, i.e.
$a^K_1=0.17\pm0.17$, $a^K_2=0.115\pm0.115$,
but keep $a^K_4=-0.015$, because the theoretical predictions are insensitive to $a_K^4$.

The twist-2 DAs for longitudinally polarized vector meson $K^*$ can
be parameterized as:
\beq
 \pksa(x) &=&  \frac{f_{K^*}}{2\sqrt{2N_c} }    6x (1-x)
    \left[1+a_{1K^*}C^{3/2}_1(t)+a_{2K^*}C^{3/2}_2(t)\right],\label{piw1b}
 \end{eqnarray}
where $f_{K^*}=200$ MeV is the decay constant of the vector meson with
longitudinal polarization, and the Gegenbauer moments are
$a_{1K^*}=0.03\pm0.03,  a_{2K^*}=0.11\pm0.11$.
As for the twist-3 DAs $\pksp$ and $\pkst$, there is no recent update
associated with those updates for twist-2 DAs, we adopt their asymptotic form:
\beq
\pksp(x) =   \frac{3f^T_{K^*}}{2\sqrt{2N_c}}(1-2x),\quad \pkst(x)
 =\frac{3f^T_{K^*}}{2\sqrt{2N_c}}(2x-1)^2,\label{piw2b}
\eeq
At last the Gegenbauer polynomials $C^{\nu}_n(t)$ are given as:
\beq
C^{1/2}_2(t)&=&\frac{1}{2}(3t^2-1), \qquad C^{1/2}_4(t)=\frac{1}{8}(3-30t^2+35t^4), \non
C^{3/2}_1(t)&=&3t, \qquad C^{3/2}_2(t)=\frac{3}{2}(5t^2-1),\non
C^{3/2}_4(t)&=&\frac{15}{8}(1-14t^2+21t^4),
\label{eq:c124}
\eeq
with $t=2x-1$.

\section{Decay amplitudes at leading order in pQCD approach}\label{ssec:lo1}

The $B \to K K^*$ decays have been studied previously in
Ref.~\cite{guo07} by using the leading order pQCD approach.
In this paper, we focus on
the calculations of some NLO contributions to these decays in the pQCD factorization approach.
For the sake of completeness, however, we firstly recalculate and present the relevant
LO decay amplitudes in this section.

\begin{figure}[tb]
\vspace{-2cm}
\centerline{\epsfxsize=19 cm \epsffile{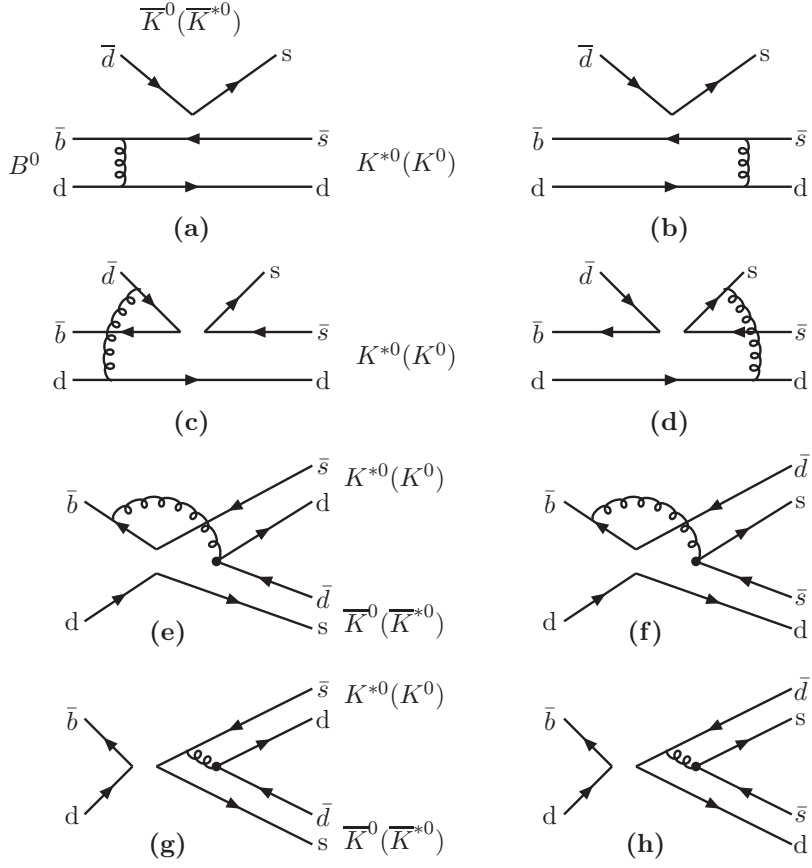}}
\vspace{-14cm}
\caption{ Diagrams contributing to the $B^0\to K^{*0}\overline{K}^{0}$ decay.
From diagram (a) and (b), the form factor $A_0^{B\to K^*}$ or $F_{0,1}^{B\to K}$
can be extracted.  Other four Feynman diagrams can be obtained by connecting
the left-hand end of the gluon line to lower $d$ quark line inside the $B^0$ meson
for (e) and (f), while to the lower $s$ or $d$ quark line for (g) and (h).}
\label{fig:fig1}
\end{figure}

\begin{figure}[t,b]
\vspace{-2cm} \centerline{\epsfxsize=19 cm \epsffile{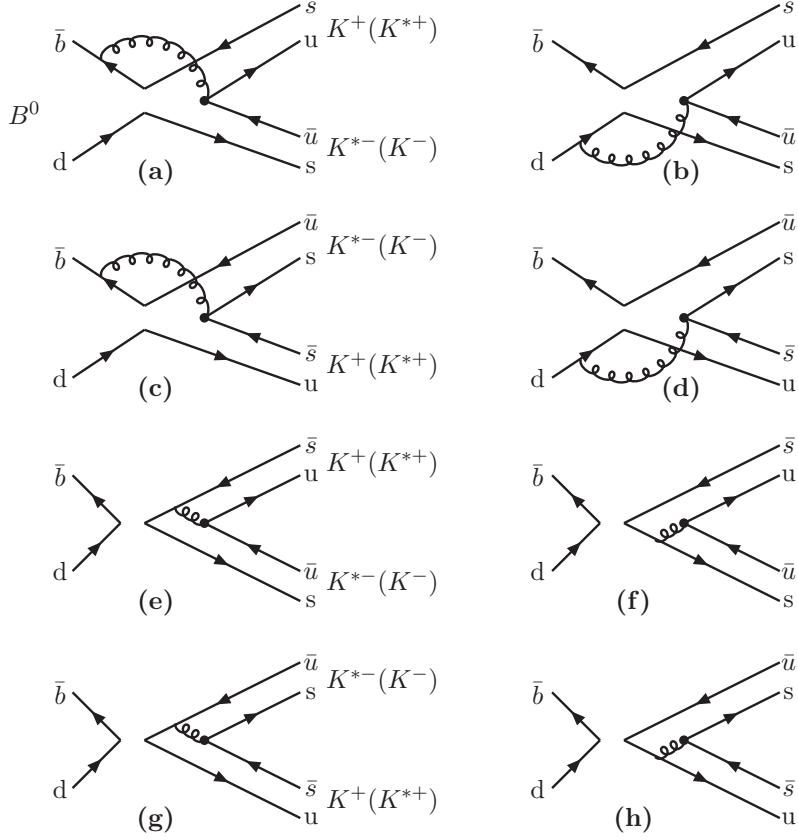}}
\vspace{-14cm}
\caption{ Diagrams contributing to the $B\to
K^+K^{*-}(K^{*+}K^-)$ decays. }
 \label{fig:fig2}
\end{figure}

\begin{figure}[t,b]
\vspace{-2cm}
\centerline{\epsfxsize=19 cm \epsffile{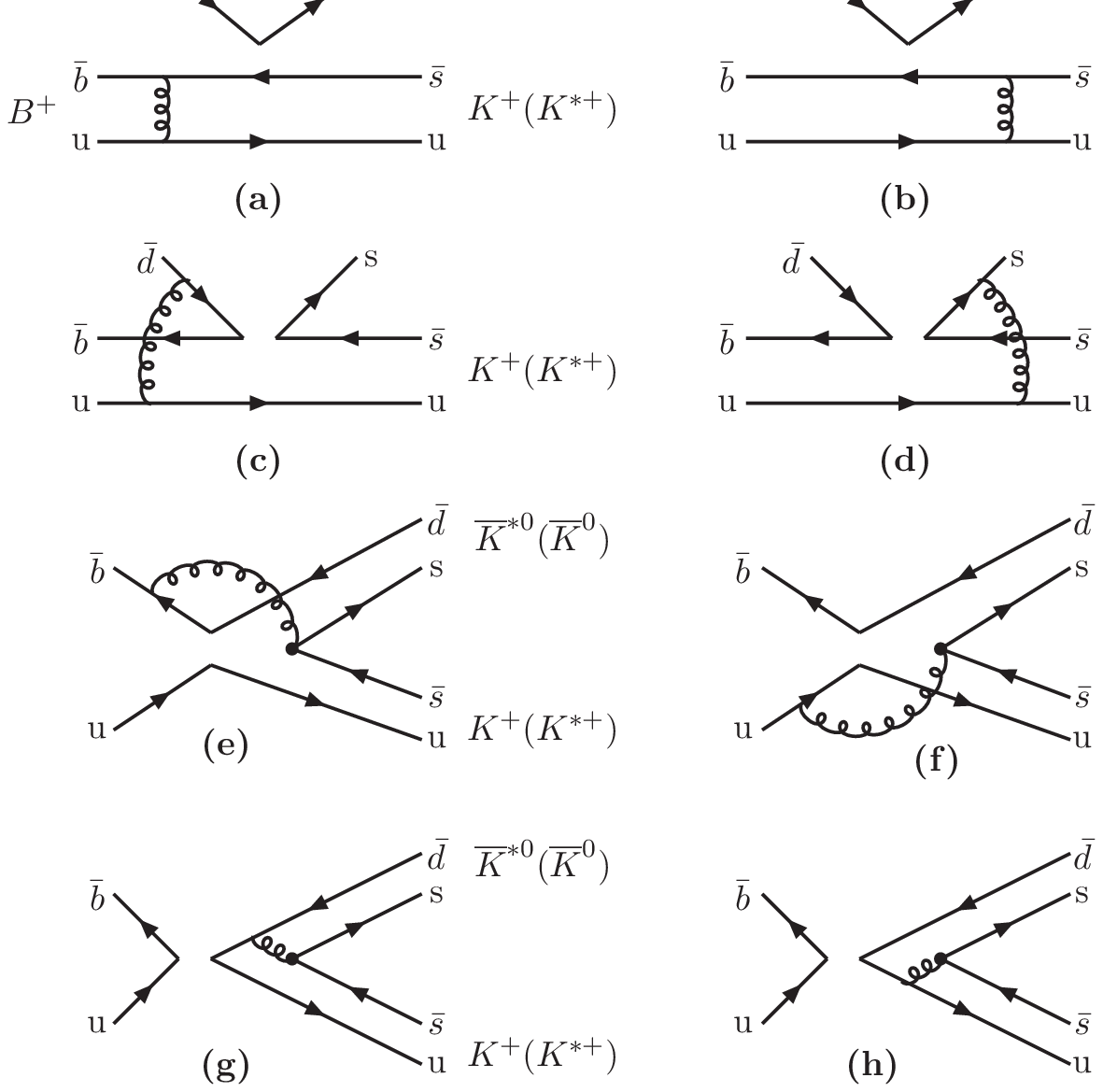}}
\vspace{-14cm}
\caption{ Diagrams contributing to the $B^+\to
K^+\overline{K}^{*0} (K^{*+}\overline{K}^{0})$ decays. }
 \label{fig:fig3}
\end{figure}

At the leading order, the relevant Feynman diagrams
for $B^0\to K^{*0}\overline{K}^{0},  K^{0}\overline{K}^{*0}$,
$B^0\to K^+ K^{*-}, K^- K^{*+}$, and $B^+ \to K^+ \overline{K}^{*0},  K^{*+}\overline{K}^{0}$
decays have been shown in Figs.~\ref{fig:fig1}, \ref{fig:fig2} and \ref{fig:fig3}.

As illustrated by Fig.~\ref{fig:fig1}, both $B^0$ and $\ov{B}^0$ can decay into
$K^{*0}\overline{K}^{0}$ and $K^0 \overline{K}^{*0}$ simultaneously. Besides of the eight
Feynman diagrams in Fig.~\ref{fig:fig1}, other four Feynman diagrams can be obtained
by connecting the left-hand end of the gluon line to the lower $d$ quark line inside
the $B^0$ meson for (e) and (f), or to the lower $s$ or $d$ quark line for (g) and (h).
For $B^0 \to K^{*0}\overline{K}^{0}$ and $K^0 \overline{K}^{*0}$ decays, only
the operators $O_{3-10}$ contribute via penguin topology. Its is a pure
penguin mode with only one kind of CKM element $\xi_t = V_{tb}^* V_{td}$, and therefore
there is no CP violation for such decays at leading order.

For $B^0 \to K^{*0}\overline{K}^{0}, K^0 \overline{K}^{*0}$ decays, we firstly consider
the case of $B \to K^{*0}$ transition where $K^{*0}$ meson
takes the spectator $d$ quark. For the $(V-A)(V-A)$ operators, the decay amplitude
corresponding to Figs.~(\ref{fig:fig1})a and (\ref{fig:fig1})b can be written as
\beq
F_{eK^*}&=& 4\sqrt2 G_F\pi C_F f_K m_B^4\int_0^1 d x_{1}
dx_{2}\, \int_{0}^{\infty} b_1 db_1 b_2 db_2\, \phi_B(x_1,b_1) \non
& & \times \left\{ \left[(1+x_2) \pksa({\xeba}) -(1-2x_2) \rks
(\pksp(\xeba) -\pkst(\xeba))\right]\right.\non
&& \left. \cdot E_e(t_a)\; h_e(x_1,x_2,b_1,b_2)
-2\rks \pksp (\xeba)\cdot E_e(t_a^{\prime})\; h_e(x_2,x_1,b_2,b_1) \right\},
\label{eq:ab-01}
\eeq
where $\rks=m_{K^*}/m_B$, $C_F=4/3$ is a color factor. The evolution factors
$E_e(t_a^{(\prime)})$ and the hard functions $h_e$  are displayed in
Appendix \ref{sec:aa}.

For the $(V-A)(V+A)$ and $(S-P)(S+P)$ operators, we find
 \beq
F_{eK^*}^{P1}&=&-F_{eK^*}, \\
F_{eK^*}^{P2}&=& 8\sqrt2 G_F\pi C_F f_K m_B^4\int_{0}^{1}d x_{1}d
x_{2}\,\int_{0}^{\infty} b_1d b_1 b_2d b_2\, \pb(x_1) \non & &
\times
 \left\{-\rk \left[ \pksa(\xeba)- \rks((2+x_2) \pksp (\xeba)+x_2\pkst(\xeba))\right]
\right.\non && \left. \cdot
E_e(t_a)\; h_e (x_1,x_2,b_1,b_2)
+ 2\rks\rk\pksp (\xeba)\; \cdot
E_e(t_a^{\prime})\; h_e(x_2,x_1,b_2,b_1) \right\} \; .
 \eeq

For the non-factorizable diagrams 1(c) and 1(d), all three meson
wave functions are involved. The decay amplitudes are
\beq
M_{eK^*}&=& \frac{16}{\sqrt{3}}G_F\pi C_F m_B^4
\int_{0}^{1}d x_{1}d x_{2}\,d x_{3}\,\int_{0}^{\infty} b_1d b_1 b_3d
b_3\, \pb(x_1,b_1) \pka(\xsba) \non
 & &\times \left\{\left[ \rks
x_2\left(\pksp(\xeba)+\pkst(\xeba)\right)+(1-x_3)\pksa(\xeba)\right
] \right. \non
 & &\left. \cdot E_e^{\prime}(t_b) h_n(x_1,x_2,1-x_3,b_1,b_3)
+E_e^{\prime}(t_b^{\prime}) h_n(x_1,x_2,x_3,b_1,b_3)\right.\non &&
 \cdot \left. \left[-(x_2+x_3)\pksa(\xeba)-\rks
x_2\left(\pksp(\xeba)-\pkst(\xsba)\right)\right]\right\}, \\
M_{eK^*}^{P1}&=&
-\frac{16}{\sqrt{3}}G_F\pi C_F m_B^4 r_K\int_{0}^{1}d x_{1}d
x_{2}\,d x_{3}\,\int_{0}^{\infty} b_1d b_1 b_3d b_3\, \pb(x_1,b_1)
\times\left\{\left[(1-x_3)\pksa(\xeba) \right.\right.\non
&&\left.\left. \cdot\left(\pkp(\xsba)-\pkt(\xsba)\right)-
\rks(1-x_3)\left(\pksp(\xeba)+\pkst(\xeba)\right)
\left(\pkp(\xsba)-\pkt(\xsba)\right)\right.\right.\non && \left.
-\rks x_2\left(\pksp(\xeba)-\pkst(\xeba)\right)
 \left(\pkp(\xsba)+\pkt(\xsba)\right)\right]E_e^{\prime}(t_b)h_n(x_1,x_2,1-x_3,b_1,b_3)
\non &&\left.  - \left[x_3\cdot
\pksa(\xeba)\left(\pkp(\xsba)+\pkt(\xsba)\right) -\rks
x_3\left(\pksp(\xeba)+\pkst(\xeba)\right) \right.\right.
\left(\pkp(\xsba)\right.\non
&&\left.\left.\left.+\pkt(\xsba)\right)-\rks
x_2\left(\pksp(\xeba)-\pkst(\xeba)\right)\left(\pkp(\xsba)-\pkt(\xsba)\right)\right]
\right.\non && \left. \times E_e^{\prime}(t_b^{\prime})
h_n(x_1,x_2,x_3,b_1,b_3) \right\},\\
M_{eK^*}^{P2}&=&0.
\eeq

For the non-factorizable annihilation diagrams (e) and (f), again all three wave functions
are involved. The decay amplitudes are
\beq
 M_{aK^*}&=&
\frac{16}{\sqrt{3}}G_F\pi C_F m_B^4\int_{0}^{1}d x_{1}d x_{2}\,d
x_{3}\,\int_{0}^{\infty} b_1d b_1 b_3d b_3\, \pb(x_1,b_1)\non &&
\times \left\{\left[(1- x_2)\pksa(\xeba) \pka(\xsba) +\rks \rk
(1-x_2) \left(\pksp(\xeba)+\pkst(\xeba) \right)
\left(\pkp(\xsba)-\right.\right.\right.\non
&&\left.\left.\left.\pkt(\xsba) \right)
 + \rks\rk x_3\left(\pksp(\xeba)-\pkst(\xeba)
\right)\left(\pkp(\xsba)+\pkt(\xsba)\right)
 \right]\right.\non &&\left.
\times E_a^{\prime}(t_c)h_{na}(x_1,x_2,x_3,b_1,b_3)
 -E_a^{\prime}(t_c^{\prime})h_{na}^{\prime}(x_1,x_2,x_3,b_1,b_3)
\left[ x_3 \pksa(\xeba) \pka(\xsba) \right.\right.\non
&&\left.\left.+4\rks \rk \pksp(\xeba)\pkp(\xsba)-
 \rks\rk(1-x_3)\left(\pksp(\xeba)+\pkst(\xeba)\right)
\left(\pkp(\xsba)\right.\right.\right.\non &&\left.\left.\left.-
 \pkt(\xsba)\right)- \rks\rk x_2
 \left(\pksp(\xeba)-\pkst(\xeba)\right)\left(\pkp(\xsba)+\pkt(\xsba)\right)\right]
 \right \}\; ,
 \eeq
\beq M_{aK^*}^{P1}&=& \frac{16}{\sqrt{3}}G_F\pi C_F
m_B^4\int_{0}^{1}d x_{1}d x_{2}\,d x_{3}\,\int_{0}^{\infty} b_1d b_1
b_3d b_3\, \pb(x_1,b_1)
 \non
& & \left\{ \left[ (1-x_2) \rks
\pka(\xsba)\left(\pksp(\xeba)+\pkst(\xeba) \right)-\rk x_3
\pksa(\xeba)\left( \pkp(\xsba) -\pkt(\xsba)\right)\right]\right.\non
&&\left.\times
E_a^{\prime}(t_c)h_{na}(x_1,x_2,x_3,b_1,b_3)-\left[-(x_2+1)\rks
\pka(\xsba)\left(\pksp(\xeba)+ \pkst(\xeba) \right)
\right.\right.\non &&\left.\left.-\rk
(x_3-2)\pksa(\xeba)\left(\pkp(\xsba)-\pkt(\xsba)\right)\right]
E_a^{\prime}(t_c^{\prime})h_{na}^{\prime}(x_1,x_2,x_3,b_1,b_3)
\right \} \; . \eeq \beq M_{aK^*}^{P2} &=&\frac{16}{\sqrt{3}}G_F\pi
C_F m_B^4\int_{0}^{1}d x_{1}d x_{2}\,d x_{3}\,\int_{0}^{\infty} b_1d
b_1 b_3d b_3\, \pb(x_1,b_1) \left\{\left[(x_2-1)\right.\right.\non
&&\left.\left.\times\pksa(\xeba)\pka(\xsba)-4\rk\rks\pksp(\xeba)\pkp(\xsba)+\rk\rks
x_2\left(\pksp(\xeba)+\pkst(\xeba)\right) \right.\right.\non
&&\left.\left. \cdot\left(\pkp(\xsba)-\pkt(\xsba)\right) +\rks\rk
(1-x_3)\left(\pksp(\xeba)-\pkst(\xeba)\right)\left(\pkp(\xsba)+\pkt(\xsba)\right)
\right] \right.\non &&\left. \cdot
E_a^{\prime}(t_e)h_{na}(x_1,x_2,x_3,b_1,b_3) +
\left[x_3\pksa(\xeba)\pka(\xsba)+x_3\rks\rk
\left(\pksp(\xeba)+\pkst(\xeba)\right)\right.\right.\quad\non
&&\left.\left.
\cdot\left(\pkp(\xsba)-\pkt(\xsba)\right)+\rks\rk(1-x_2)\left(\pksp(\xeba)-\pkst(\xeba)\right)
\left(\pkp(\xsba)+\pkt(\xsba)\right) \right]\right.\non && \left.
\times
E_a^{\prime}(t_e^{\prime})h_{na}^{\prime}(x_1,x_2,x_3,b_1,b_3)\right\}.
 \eeq

The factorizable annihilation diagrams (g) and (h) involve only
$K^*$ and $K$ wave functions. There are also three kinds of decay
amplitudes for these diagrams.
$F_{aK^*}$ is for $(V-A)(V-A)$
\beq
F_{aK^*}&=&F_{aK^*}^{P1}= 4\sqrt{2}G_F\pi C_F f_B m_B^4
\int_{0}^{1}dx_{2}\,d x_{3}\, \int_{0}^{\infty} b_2d b_2b_3d b_3 \,
\left\{-\left[ (1-x_2) \pksa(\xeba) \right.\right.\non
&&\left.\left.\cdot\pka(\xsba) +4 \rk
\rks\pksp(\xeba)\pkp(\xsba)-2\rks\rk
x_2\pkp(\xsba)\left(\pksp(\xeba)+\pkst(\xeba)\right)\right]
\right.\non && \left.\cdot E_a(t_d)h_a(x_3,1-x_2,b_3,b_2)+ \left[x_3
\pksa(\xeba) \pka(\xsba)+2 \rk \rks \pksp(\xeba)\right.\right.\non
&&\left.\left.\cdot \left(\pkp(\xsba)+\pkt(\xsba)\right)+2\rk\rks
x_3\pksp(\xeba)\left(\pkp(\xsba)-\pkt(\xsba)\right)\right]\right.\non
&& \left. \times E_a(t_d^{\prime}) h_a(1-x_2,x_3,b_2,b_3) \right
\}\; , \eeq \beq F_{aK^*}^{P2}&=& -8\sqrt{2}G_F\pi C_F f_B m_B^4
\int_{0}^{1}d x_{2}\,d x_{3}\,\int_{0}^{\infty} b_2d b_2b_3d b_3
\,\non &&\times \left\{ \left[\rks(1-x_2)
\left(\pksp(\xeba)-\pkst(\xeba)\right)\pka(\xsba)+2\rk \pksa(\xeba)
\pkp(\xsba) \right]\right.
 \non
&&\left.\times E_a(t_d) h_a(x_3,1-x_2,b_3,b_2)\right.
 \non
&&\left.+\left[2\rks
\pksp(\xeba)\pka(\xsba)+x_3\rk\pksa(\xeba)(\pkp(\xsba)+\pkt(\xsba))\right]
\right.\non &&\left.\times E_a(t_d^{\prime})
h_a(1-x_2,x_3,b_2,b_3)\right\}\; \label{eq:mapip2}.
 \eeq

For the case of $B^0 \to K^0$ transition where $K^0$ meson
takes up the spectator $d$ quark, as shown in Fig.~\ref{fig:fig1},
it is straightforward to find the decay amplitudes by following
the same procedure as the case of $B^0 \to K^{*0}$ transition.
\beq
F_{eK}&=& 16\pi C_F m_B^2\int_0^1 d x_{1} dx_{2}\, \int_{0}^{\infty} b_1 db_1 b_2
db_2\, \phi_B(x_1,b_1) \non & & \times \left\{ \left[(1+x_2)
\pka({\xeba}) +(1-2x_2) \rk (\pkp(\xeba) -\pkt(\xeba))\right]
E_e(t_a)h_e(x_1,x_2,b_1,b_2)\right.\non && \left. +2\rk \pkp (\xeba)
E_e(t_a^{\prime})h_e(x_2,x_1,b_2,b_1) \right\},
\label{eq:ab1}\\
F_{eK}^{P1}&=&F_{eK}, \quad F_{eK}^{P2}=0.
\eeq
\beq
M_{eK}&=& \frac{16}{\sqrt{3}}G_F\pi C_F m_B^4
\int_{0}^{1}d x_{1}d x_{2}\,d x_{3}\,\int_{0}^{\infty} b_1d b_1 b_3d
b_3\, \pb(x_1,b_1) \pksa(\xsba) \non
 & &\times \left\{\left[
-\rk
x_2\left(\pkp(\xeba)+\pkt(\xeba)\right)+(1-x_3)\pka(\xeba)\right ]
\right. \non
 & &\left. \cdot E_e^{\prime}(t_b) h_n(x_1,x_2,1-x_3,b_1,b_3)
+E_e^{\prime}(t_b^{\prime}) h_n(x_1,x_2,x_3,b_1,b_3)\right.\non &&
 \cdot \left. \left[-(x_2+x_3)\pka(\xeba)+\rk
x_2\left(\pkp(\xeba)-\pkt(\xsba)\right)\right]\right\}, \\
M_{eK}^{P1}&=&
-\frac{16}{\sqrt{3}}G_F\pi C_F m_B^4 \rks\int_{0}^{1}d x_{1}d
x_{2}\,d x_{3}\,\int_{0}^{\infty} b_1d b_1 b_3d b_3\, \pb(x_1,b_1)
\times\left\{\left[(1-x_3)\pka(\xeba) \right.\right.\non
&&\left.\left. \cdot\left(\pksp(\xsba)-\pkst(\xsba)\right)+
\rk(1-x_3)\left(\pkp(\xeba)+\pkt(\xeba)\right)
\left(\pksp(\xsba)-\pkst(\xsba)\right)\right.\right.\non && \left.
+\rk x_2\left(\pkp(\xeba)-\pkt(\xeba)\right)
 \left(\pksp(\xsba)+\pkst(\xsba)\right)\right]E_e^{\prime}(t_b)h_n(x_1,x_2,1-x_3,b_1,b_3)
\non &&\left.  - \left[x_3\cdot
\pka(\xeba)\left(\pksp(\xsba)+\pkst(\xsba)\right) +\rk
x_3\left(\pkp(\xeba)+\pkt(\xeba)\right) \right.\right.
\left(\pksp(\xsba)\right.\non
&&\left.\left.\left.+\pkst(\xsba)\right)+\rk
x_2\left(\pkp(\xeba)-\pkt(\xeba)\right)\left(\pksp(\xsba)-\pkst(\xsba)\right)\right]
\right.\non && \left. \times E_e^{\prime}(t_b^{\prime})
h_n(x_1,x_2,x_3,b_1,b_3) \right\}.
\eeq
\beq
 M_{aK}&=& \frac{16}{\sqrt{3}}G_F\pi C_F m_B^4\int_{0}^{1}d x_{1}d x_{2}\,d
x_{3}\,\int_{0}^{\infty} b_1d b_1 b_3d b_3\, \pb(x_1,b_1)\non &&
\times \left\{\left[(1- x_2)\pka(\xeba) \pksa(\xsba) -\rk \rks
(1-x_2) \left(\pkp(\xeba)+\pkt(\xeba) \right)
\left(\pksp(\xsba)-\right.\right.\right.\non
&&\left.\left.\left.\pkst(\xsba) \right)
 -\rk\rks x_3\left(\pkp(\xeba)-\pkt(\xeba)
\right)\left(\pksp(\xsba)+\pkst(\xsba)\right)
 \right]\right.\non &&\left.
\times E_a^{\prime}(t_c)h_{na}(x_1,x_2,x_3,b_1,b_3)
 -E_a^{\prime}(t_c^{\prime})h_{na}^{\prime}(x_1,x_2,x_3,b_1,b_3)
\left[ x_3 \pka(\xeba) \pksa(\xsba) \right.\right.\non
&&\left.\left.-4\rk \rks \pkp(\xeba)\pksp(\xsba)-
 \rk\rks(1-x_3)\left(\pkp(\xeba)+\pkt(\xeba)\right)
\left(\pksp(\xsba)\right.\right.\right.\non &&\left.\left.\left.-
 \pkst(\xsba)\right)+ \rk\rks x_2
 \left(\pkp(\xeba)-\pkt(\xeba)\right)\left(\pksp(\xsba)+\pkst(\xsba)\right)\right]
 \right \}\; ,
\eeq
\beq M_{aK}^{P1}&=& \frac{16}{\sqrt{3}}G_F\pi C_F m_B^4\int_{0}^{1}d
x_{1}d x_{2}\,d x_{3}\,\int_{0}^{\infty} b_1d b_1 b_3d b_3\,\pb(x_1,b_1)
 \non
& & \left\{ \left[ (1-x_2) \rks
\pksa(\xsba)\left(\pkp(\xeba)+\pkt(\xeba) \right)-\rks x_3
\pka(\xeba)\left( \pksp(\xsba)
-\pkst(\xsba)\right)\right]\right.\non &&\left.\times
E_a^{\prime}(t_c)h_{na}(x_1,x_2,x_3,b_1,b_3)-\left[(x_2+1)\rk
\pksa(\xsba)\left(\pkp(\xeba)+ \pkt(\xeba) \right)
\right.\right.\non &&\left.\left.-\rks
(x_3-2)\pka(\xeba)\left(\pksp(\xsba)-\pkst(\xsba)\right)\right]
E_a^{\prime}(t_c^{\prime})h_{na}^{\prime}(x_1,x_2,x_3,b_1,b_3)
\right \} \; ,
\eeq
\beq M_{aK}^{P2} &=&\frac{16}{\sqrt{3}}G_F\pi
C_F m_B^4\int_{0}^{1}d x_{1}d x_{2}\,d x_{3}\,\int_{0}^{\infty} b_1d
b_1 b_3d b_3\, \pb(x_1,b_1) \left\{\left[(x_2-1)\right.\right.\non
&&\left.\left.\times\pka(\xeba)\pksa(\xsba)+4\rks\rk\pkp(\xeba)\pksp(\xsba)-\rks\rk
x_2\left(\pkp(\xeba)+\pkt(\xeba)\right) \right.\right.\non
&&\left.\left. \cdot\left(\pksp(\xsba)-\pkst(\xsba)\right) -\rk\rks
(1-x_3)\left(\pkp(\xeba)-\pkt(\xeba)\right)\left(\pksp(\xsba)+\pkst(\xsba)\right)
\right] \right.\non &&\left. \cdot
E_a^{\prime}(t_e)h_{na}(x_1,x_2,x_3,b_1,b_3) +
\left[x_3\pka(\xeba)\pksa(\xsba)-x_3\rk\rks\left(\pkp(\xeba)+\pkt(\xeba)\right)\right.\right.\quad\non
&&\left.\left.
\cdot\left(\pksp(\xsba)-\pkst(\xsba)\right)-\rk\rks(1-x_2)\left(\pkp(\xeba)-\pkt(\xeba)\right)
\left(\pksp(\xsba)+\pkst(\xsba)\right) \right]\right.\non && \left.
\times
E_a^{\prime}(t_e^{\prime})h_{na}^{\prime}(x_1,x_2,x_3,b_1,b_3)\right\}.
\eeq

\beq
F_{aK}&=&F_{aK}^{P1}= 4\sqrt{2}G_F\pi C_F f_B m_B^4
\int_{0}^{1}dx_{2}\,d x_{3}\, \int_{0}^{\infty} b_2d b_2b_3d b_3 \,
\left\{-\left[ (1-x_2) \pka(\xeba) \right.\right.\non
&&\left.\left.\cdot\pksa(\xsba) -4 \rks
\rk\pkp(\xeba)\pksp(\xsba)+2\rk\rks
x_2\pksp(\xsba)\left(\pkp(\xeba)+\pkt(\xeba)\right)\right]
\right.\non && \left.\cdot E_a(t_d)h_a(x_3,1-x_2,b_3,b_2)+ \left[x_3
\pka(\xeba) \pksa(\xsba)-2 \rks \rk \pkp(\xeba)\right.\right.\non
&&\left.\left.\cdot \left(\pksp(\xsba)+\pkst(\xsba)\right)-2\rks\rk
x_3\pkp(\xeba)\left(\pksp(\xsba)-\pkst(\xsba)\right)\right]\right.\non
&& \left. \times E_a(t_d^{\prime}) h_a(1-x_2,x_3,b_2,b_3) \right
\}\; , \eeq \beq F_{aK}^{P2}&=& 8\sqrt{2}G_F\pi C_F f_B m_B^4
\int_{0}^{1}d x_{2}\,d x_{3}\,\int_{0}^{\infty} b_2d b_2b_3d b_3
\,\non &&\times \left\{ \left[\rk(1-x_2)
\left(\pkp(\xeba)-\pkt(\xeba)\right)\pksa(\xsba)-2\rks \pka(\xeba)
\pksp(\xsba) \right]\right.
 \non
&&\left.\times E_a(t_d) h_a(x_3,1-x_2,b_3,b_2)\right.
 \non
&&\left.+\left[2\rk
\pkp(\xeba)\pksa(\xsba)-x_3\rks\pka(\xeba)(\pksp(\xsba)+\pkst(\xsba))\right]
\right.\non &&\left.\times E_a(t_d^{\prime})
h_a(1-x_2,x_3,b_2,b_3)\right\}\; .
\label{eq:cd-05}
\eeq

Combining the contributions from different diagrams in Fig.~\ref{fig:fig1},
the total decay amplitude for $B^0\to K^{*0}\overline{K}^{0}$ and $K^{0}\overline{K}^{*0}$
decay can be written as
\beq
{\cal M}(B^0 & \to &  K^{*0}\overline{K}^{0} + K^{0}\overline{K}^{*0} ) =
-\xi_t \left \{ \left ( F_{eK} + F_{eK}^* \right ) \left( a_4 -\frac{1}{2} a_{10}\right)
+ F_{eK^*}^{P_2}\left(a_6-\frac{1}{2} a_8 \right)\right.\non
&&\left.
 +\left ( M_{eK} + M_{eK}^*\right )\left(C_3-\frac{1}{2}C_9\right)
 + \left ( M_{eK}^{P_1}+ M_{eK^*}^{P_1} \right )\left(C_5-\frac{1}{2}C_7\right)
\right. \non
&&\left.
+ \left ( M_{aK} +  M_{aK^*}\right ) \left(C_3+2C_4-\frac{1}{2}C_9-C_{10}\right)
\right.\non
&&
\left.
+ \left ( M_{aK}^{P_1}+ M_{aK^*}^{P_1} \right )\left(C_5-\frac{1}{2}C_7\right)
+\left ( M_{aK}^{P_2}+ M_{aK^*}^{P_2}\right ) \left( 2 C_6- C_8\right)
\right.
 \non
 &&\left.
+F_{aK} \left(2 a_3 + a_4 + 2a_5 -a_7 - a_9 -\frac{1}{2} a_{10}) \right)
\right.
 \non
 &&\left.
+ F_{aK^*} \left(2 a_3 + a_4 - a_9 -\frac{1}{2} a_{10}) \right)
+\left ( F_{aK}^{P_2}+ F_{aK^*}^{P_2} \right)\left(a_6-\frac{1}{2}a_8 \right)
\right \}
\label{eq:m001}
\eeq

For $B^0\to K^+K^{*-}(K^{*+}K^-)$ decays£¬ as illustrated in Fig.\ref{fig:fig2},
only annihilation diagrams contribute at leading order. Again, both $B^0$ and $\ov{B}^0$
mesons can decay into  the final state $K^+K^{*-}$ and its charge-conjugate
state $K^- K^{*+}$.
For $B^+ \to K^+\overline{K}^{*0}$ and $K^{*+}\overline{K}^{0}$ decays, as shown in
Fig.~\ref{fig:fig3}, the factorizable emission diagram, the non-factorizable
spectator and annihilation diagrams contribute simultaneously.

Following the same procedure as for $B^0\to  K^{*0}\overline{K}^{0}/ K^{0}\overline{K}^{*0}$
decays, we find the total decay amplitude for the later two decay modes:
 \beq
 {\cal M}(B^0 &\to & K^+ K^{*-} +  K^- K^{*+} ) =
 \xi_u \left[  \left( M_{aK} + M_{aK^*} \right ] C_2 +
 \left ( F_{aK}+ F_{aK^*}\right) a_2 \right]\non
&&
 -\xi_t \left \{ \left ( M_{aK} + M_{aK^*} \right )\left(2C_4+\frac{1}{2}C_{10}\right)
 + \left ( M_{aK}^{P_2} +  M_{aK^*}^{P_2} \right ) \left(2C_6 + \frac{1}{2} C_{8}\right)
 \right.
 \non
& &\left.
+\left ( F_{aK} + F_{aK^*} \right)
\left(2 a_3+ 2a_5 + \frac{1}{2} a_7 + \frac{1}{2} a_9 \right) \right\}.
\label{eq:m4}
 \eeq

 \beq
 {\cal M}(B^+ \to K^+ \ov K^{0*}) &=& \xi_u \left( M_{aK}\; C_1 + F_{aK}\; a_1 \right)
 -\xi_t \left \{ F_{eK}\left( a_4-\frac{1}{2} a_{10}\right)
 \right. \non
& &\left.
+M_{eK}\left(C_3-\frac{1}{2}C_9\right)+M_{eK}^{P_1}\left(C_5-\frac{1}{2}C_7\right)
+M_{aK}\left(C_3+C_9\right) \right.
 \non
&&\left.+M_{aK}^{P_1}\left(C_5+C_7\right)
+F_{aK}\left( a_4 + a_{10} \right)
+F_{aK}^{P_2}\left( a_6 + a_8\right)  \right \} ,
\label{eq:m5}
 \eeq
 \beq
 {\cal M}(B^+ \to K^{*+} \ov K^0) &=&
 \xi_u \left( M_{aK^*}\; C_1 + F_{aK^*}\; a_1 \right)\non
 && -\xi_t \left \{ F_{eK^*}\left(a_4-\frac{1}{2}\; a_{10}\right)
+F_{eK^*}^{P_2}\left( a_6-\frac{1}{2}\; a_8\right)
\right.\non
&& \left.
+ M_{eK^*}\left( C_3-\frac{1}{2}C_9\right)
+ M_{eK^*}^{P_1}\left(C_5 -\frac{1}{2}C_7\right)
\right. \non
 &&\left.
+ M_{aK^*}\left(C_3+C_9\right)
+ M_{aK^*}^{P_1}\left(C_5+C_7\right)
\right. \non
&&\left.
+ F_{aK^*} \left(a_4 + a_{10}\right)
+ F_{aK^*}^{P_2}\left(a_6 + a_8\right) \right \} ,
\label{eq:m6}
 \eeq

In the decay amplitudes of Eqs.(\ref{eq:m001}) - (\ref{eq:m6}), the coefficients $a_i$,
the standard combination of the Wilson coefficients $C_i$,
have been defined as usual
\beq
a_1&=& C_2 + \frac{C_1}{3}, \quad a_2= C_1 + \frac{C_2}{3}, \non
a_{i}&=& C_i + \frac{C_{i+1}}{3}, \ \ {\rm for} \ \ i=3,5,7,9,\non
a_{i}&=& C_i + \frac{C_{i-1}}{3}, \ \ {\rm for} \ \ i=4,6,8,10.
\label{eq:aimu}
\eeq

Based on these decay amplitudes, the leading order pQCD predictions for the branching
ratios and CP violating asymmetries of the considered decays can be
calculated \cite{guo07}.


\section{NLO contributions to $B\to KK^*$ decays in pQCD}

The power counting in the pQCD factorization approach \cite{nlo05} is different
from that in the QCD factorization\cite{bbns99}.
When compared with the previous LO calculations
in pQCD \cite{guo07}, the following NLO contributions will be included:

\begin{enumerate}

\item
The LO Wilson coefficients $C_i(\mw)$ will be replaced by those at NLO level in NDR scheme
\cite{buras96}. As mentioned in last section,  the strong coupling constant
$\alpha_s(t)$ at two-loop level as given in Eq.~(\ref{eq:asnlo}), and the NLO RG evolution
matrix $U(t,m,\alpha)$, as defined in Ref.~\cite{buras96}, will be used here:
\beq
U(m_1,m_2,\alpha) = U(m_1,m_2) + \frac{\alpha}{4\pi} R(m_1,m_2)
\label{eq:um1m2a}
\eeq
where the function  $U(m_1,m_2)$ and $R(m_1,m_2)$ represent the QCD and QED evolution and
have been defined in Eq.~(6.24) and (7.22) in Ref.~\cite{buras96}. We also introduce a cut-off
 $\Lambda_{cut} = 1$ GeV for low energy scale in the final integration.

\item The NLO contributions to the hard kernel $H$, including the
vertex corrections, the quark loops, and the magnetic penguin \cite{nlo05}.

\end{enumerate}

\subsection{Vertex corrections}

\begin{figure}[tb]
\vspace{-3cm}
\centerline{\epsfxsize=19 cm \epsffile{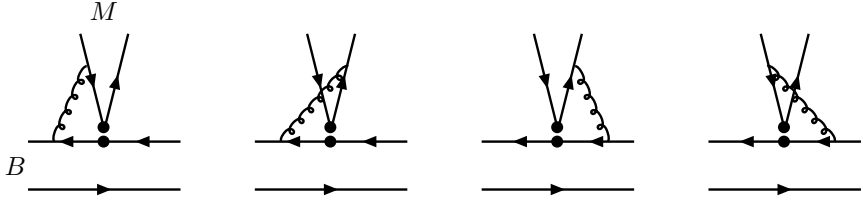}}
\vspace{-21cm}
\caption{NLO vertex corrections to the factorizable amplitudes. }
\label{fig:vc}
\end{figure}

The vertex corrections to the factorizable emission diagrams, as illustrated by
Fig.~(\ref{fig:fig2}), have been calculated years ago in the QCD factorization
appeoach\cite{bbns99,npb675}.
According to Ref.~\cite{nlo05}, the difference of the calculations induced by
considering or not considering the parton transverse momentum is rather small, say
less than $10\%$, and therefore can be neglected.
Consequently, one can use the vertex corrections as given in Ref.~\cite{npb675} directly.
The vertex corrections can be absorbed into the re-definition of the Wilson coefficients
$a_i(\mu)$ by adding a vertex-function $V_i(M)$ to them
\cite{bbns99,npb675}
\beq
a_i(\mu)&\to & a_i(\mu) +\frac{\alpha_s(\mu)}{4\pi}C_F\frac{C_i(\mu)}{3} V_i(M),\ \ for
\ \ i=1,2; \non
a_j(\mu)&\to & a_j(\mu)+\frac{\alpha_s(\mu)}{4\pi}C_F\frac{C_{j\pm
1}(\mu)}{N_c} V_j(M), \ \  for \ \ j=3-10,
\label{eq:aimu-2}
\eeq
where M is the meson emitted from the weak vertex. When $M$ is a
pseudo-scalar meson, the vertex functions $V_{i}(M)$ are given ( in the NDR
scheme) in Refs.~\cite{nlo05,npb675}:
\beq
V_i(M)&=&\left\{ \begin{array}{cc}
12\ln\frac{m_b}{\mu}-18+\frac{2\sqrt{2N_c}}{f_M}\int_{0}^{1}dx\phi_M^A(x)g(x),
& {\rm for}\quad i= 1-4,9,10,\\
-12\ln\frac{m_b}{\mu}+6-\frac{2\sqrt{2N_c}}{f_M}\int_{0}^{1}dx\phi_M^A(x)g(1-x),
&{\rm for}\quad i= 5,7,\\
-6+\frac{2\sqrt{2N_c}}{f_M}\int_{0}^{1}dx\phi_M^P(x)h(x),
&{\rm for} \quad i= 6,8,\\
\end{array}\right.
\label{eq:vim}
\eeq
where $f_M$ is the decay constant of the meson M; $\phi_M^A(x)$ and $\phi_M^P(x)$ are
the twist-2 and twist-3 distribution amplitude of the meson M, respectively.
For a vector meson M, $\phi_M^A(\phi_M^P)$ is replaced by $\phi_M (\phi_M^s)$
and $f_M$ by $f_M^T$ in the third line of the above formulas. The hard-scattering
functions $g(x)$ and $h(x)$ in Eq.~(\ref{eq:vim}) are:
\beq
g(x)&=& 3\left(\frac{1-2x}{1-x}\ln x-i\pi\right)\non
&& +\left[2Li_2(x)-\ln^2x+\frac{2\ln
x}{1-x}-\left(3+2i\pi\right)\ln x -(x\leftrightarrow 1-x)\right],\\
h(x)&=&2Li_2(x)-\ln^2x-(1+2i\pi)\ln x-(x\leftrightarrow 1-x),
\eeq
where $Li_2(x)$ is the dilogarithm function. As shown in Ref.~\cite{nlo05},
the $\mu$-dependence of the Wilson coefficients $a_i(\mu)$ will be improved generally
by the inclusion of the vertex corrections.


\subsection{Quark loops}

\begin{figure}[tb]
\vspace{-4cm}
\centerline{\epsfxsize=19 cm \epsffile{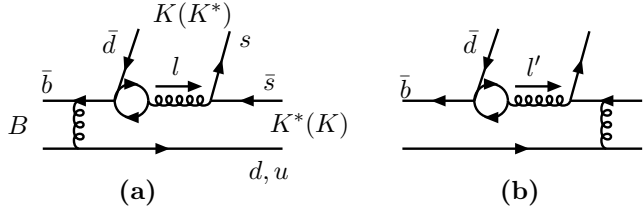}}
\vspace{-20cm}
\caption{Quark-loop diagrams contributing to $B^0\to K^{*0} \ov{K}^0 + K^{0} \ov{K}^{*0}$,
$B^+\to K^{*+} \ov{K}^0$ and $K^{+} \ov{K}^{*0}$ decays.}
\label{fig:qloops}
\end{figure}

The contribution from the so-called ``quark-loops" is a kind of penguin correction
with the four quark operators insertion, as illustrated by Fig.~(\ref{fig:qloops}).
In fact this is generally called BSS mechanism\cite{bss79}, which
plays a very important role in CP violation. We here include quark-loop amplitude from the
operators $O_{1,2}$ and $O_{3-6}$ only. The quark loops from $O_{7-10}$
will be neglected due to their smallness.

For the $b\to d$ transition, the contributions from the various
quark loops are described by the effective Hamiltonian $H_{eff}^{(q)}$ \cite{nlo05},
\beq
H_{eff}^{(q)}&=&-\sum\limits_{q=u,c,t}\sum\limits_{q{\prime}}\frac{G_F}{\sqrt{2}}
V_{qb}V_{qd}^{*}\frac{\alpha_s(\mu)}{2\pi}\; C^{(q)}(\mu,l^2)\; \left(\ov{d}\gamma_\rho
\left(1-\gamma_5\right)T^ab\right)\left(\ov{q}^{\prime}\gamma^\rho
T^a q^{\prime}\right),
\eeq
where $l^2$ being the invariant mass of the gluon, which connects the quark loops with the
$\ov{q}' q$ pair as shown in Fig.~\ref{fig:qloops}.
The functions $C^{(q)}(\mu,l^2)$ can be written as
\beq
C^{(q)}(\mu,l^2)&=&\left[G^{(q)}(\mu,l^2)-\frac{2}{3}\right]C_2(\mu),
\label{eq:qlc}
\eeq
for $q=u,c$ and
\beq
C^{(t)}(\mu,l^2)&=&\left[G^{(s)}(\mu,l^2)-\frac{2}{3}\right]
C_3(\mu)+\sum\limits_{q{\prime\prime}=u,d,s,c}G^{(q^{\prime\prime})}(\mu,l^2)
\left[C_4(\mu)+C_6(\mu)\right].\label{eq:eh}
\eeq
The integration function $G^{(q)}(\mu,l^2)$ for the loop of the
quarks $q=(u, d, s, c)$ is defined as \cite{nlo05}
\beq
G^{(q)}(\mu,l^2)&=&-4\int_{0}^{1}dx\; x(1-x)\ln \frac{m_q^2-x(1-x)l^2}{\mu^2},
\eeq
where $m_q$ is the quark mass. The explicit  expressions of the function
$G^{(q)}(\mu,l^2)$ after the integration can be found, for example, in
Ref.~\cite{nlo05}.

It is straightforward to calculate the decay amplitude for Fig.(!\ref{fig:qloops})a and
(\ref{fig:qloops})b. For the case of $B \to K^* $ or $B \to K$ transition, we find the
corresponding decay amplitude $M^{(q)}_{K^*K}$ and $M^{(q)}_{K K^*}$ with $q=u,c,t$,
respectively;
\beq
M^{(q)}_{K^*K}&=&-\frac{4}{\sqrt{3}}G_F
C_F^2m_B^4\int_{0}^{1}d x_{1}d x_{2}\,d x_{3}\,\int_{0}^{\infty}
b_1d b_1 b_2d b_2\,\phi_B(x_1,b_1)\non
&& \cdot \left\{ \left\{\left(1+x_2\right)
 \pksa(\xeba)\pka(\xsba)\right.\right.\non && \left.\left.
-\rks\left(1-2x_2\right)\left[ \pksp(\xeba)-\pkst(\xeba)\right ]\pka(\xsba)-
2\rk\pksa(\xeba)\pkp(\xsba)\right.\right.\non&&\left.\left.
+ 2\rks\rk\left[  \left( 2 +x_2\right)\pksp(\xeba)+
x_2\pkst(\xeba)\right ]
\pkp(\xsba)\right\}\right.\non
&& \left.
\hspace{1cm}\cdot E^{(q)}(t_q,l^2)h_e(x_2,x_1,b_2,b_1)
\right.\non && \left.
+ \left\{-2\rks \pksp(\xeba)\pka(\xsba)+4\rks\rk\pksp(\xeba)\pkp(\xsba)\right \}\right.\non
&& \left.
\hspace{1cm} \cdot E^{(q)}(t^{\prime}_q,l^{\prime
2}) h_e(x_1,x_2,b_1,b_2)\right\}\label{eq:qlmp}
\eeq
and
\beq
M^{(q)}_{KK^*}&=&-\frac{4}{\sqrt{3}}G_F C_F^2m_B^4\int_{0}^{1}d
x_{1}d x_{2}\,d x_{3}\,\int_{0}^{\infty} b_1d b_1 b_2d
b_2\,\phi_B(x_1,b_1)\non
&& \cdot \left\{ \left\{\left(1+x_2\right)
 \pka(\xeba)\pksa(\xsba)\right.\right.\non && \left.\left.
+\rk\left(1-2x_2\right)\left [\pkp(\xeba)-\pkt(\xeba)\right ]\pksa(\xsba)
- 2\rks\pka(\xeba)\pksp(\xsba)
\right.\right.\non&&\left.\left.
-2\rk\rks\left[ \left(2 +x_2\right)\pkp(\xeba)+
x_2\pkt(\xeba)\right]\pksp(\xsba)\right \}\right.\non
&& \left.
\hspace{1cm}\cdot E^{(q)}(t_q,l^2)h_e(x_2,x_1,b_2,b_1)
\right.\non && \left.
+\left\{2\rk \pkp(\xeba)\pksa(\xsba)-4\rk\rks\pkp(\xeba)\pksp(\xsba)\right\}\right.\non
&& \left.
\hspace{1cm} \cdot E^{(q)}(t^{\prime}_q,l^{\prime 2}) h_e(x_1,x_2,b_1,b_2)\right\}, \ \
\eeq
where $r_K =m_0^K/m_B, r_{K^*}=m_{K^*}/m_B$, the evolution factors take the form of
\beq
E^{(q)}(t,l^2)&=& C^{(q)}(t,l^2)\;\alpha_s^2(t)\cdot \exp\left [ -S_{ab} \right ],
\eeq
with the Sudakov factor $S_{ab}$ and the hard function $h_e(x_1,x_2,b_1,b_2) $
as given in Eq.~(\ref{eq:sab}) and Eq.~(\ref{eq:he}) respectively, and finally
the hard scales and the gluon invariant masses are
\beq
t_{q}&=& {\rm max}(\sqrt{x_2}m_B,\sqrt{x_1x_2}m_B,\sqrt{(1-x_2) x_3}m_B,1/b_1,1/b_2);,\non
t_{q}^{\prime}&=& {\rm max}(\sqrt{x_1}m_B,\sqrt{x_1 x_2}m_B,\sqrt{|x_3-x_1|}m_B,1/b_1,1/b_2),
\label{eq:tq-tqp}\\
l^2          &=& (1-x_2) x_3 m_B^2 - |\bfkk_{\rm 2T} -\bfkk_{\rm 3T} |^2
\approx (1-x_2) x_3 m_B^2 , \non
l^{\prime 2} &=& (x_3-x_1) m_B^2- |\bfkk_{\rm 1T} -\bfkk_{\rm 3T} |^2
\approx (x_3-x_1) m_B^2.
\eeq

Finally, the total ``quark-loop" contribution to the considered $B \to KK^*$ decays can be written as
\beq
M_{KK^*}^{(ql)} &=&  <K^* K|{\cal H}_{eff}^{q}|B>
= \sum_{q=u,c,t} \lambda_q \; \left [ M^{(q)}_{K^*K} + M^{(q)}_{KK^*} \right],
\eeq
where $\lambda_q = V_{qb}V_{qd}^{*}$.

From the functions $C^{(q)}(\mu,l^2)$, one can see that the quark-loop amplitudes depend
on both the renormalization scale $\mu$ and the gluon invariant mass $l^2$.
In the naive factorization approach, the assumption of a constant $l^2$, $l^2\sim m_b^2/2$,
introduces a large theoretical uncertainty as making predictions.
In the pQCD approach, however, $l^2$ is related to the parton
momenta unambiguously.
Because of the absence of the end-point singularities associated with $l^2, l^{\prime}\to 0$,
in Fig.~(\ref{fig:qloops})a and (\ref{fig:qloops})b respectively,
we have dropped the parton transverse momenta $k_T$ in
$l^2,l^{\prime 2}$ for simplicity \cite{nlo05}.

From Fig.~(\ref{fig:qloops}), it is easy to see that the "quark-loop" diagrams
contribute only to $B^0\to K^{*0} \ov{K}^0 + K^{0} \ov{K}^{*0}$
and  $B^+\to K^{*+} \ov{K}^0, K^{+} \ov{K}^{*0}$ decays. For $B^0 \to K^{+} K^{*-}$ and
$K^{*+} K^{-}$ decays, there is no such kind of NLO contributions.

\subsection{Chromo-magnetic penguin contributions}

\begin{figure}[tb]
\vspace{-4cm}
\centerline{\epsfxsize=19cm \epsffile{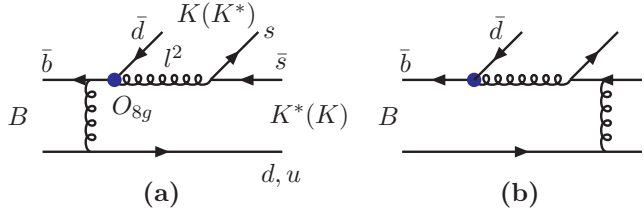}}
\vspace{-19cm}
\caption{Chromo-magnetic penguin ($O_{8g}$) diagrams contributing to
$B^0\to K^{*0} \ov{K}^0 + K^{0} \ov{K}^{*0}$,
$B^+\to K^{*+} \ov{K}^0$ and $K^{+} \ov{K}^{*0}$ decays.}
\label{fig:mp}
\end{figure}

As illustrated by Fig.~(\ref{fig:mp}), the chromo-magnetic penguin operator $O_{8g}$ also
contribute to $B \to K K^*$ decays at NLO level.
The corresponding weak effective Hamiltonian contains the $b\to d g$ transition,
\beq
{\cal H}_{eff}^{cmp}&=&-\frac{G_F}{\sqrt{2}}V_{tb}V_{td}^* \; C_{8g}^{eff} \; O_{8g},
\label{eq:heffo8g}
\eeq
with the chromo-magnetic penguin operator,
\beq
O_{8g}&=&\frac{g_s}{8\pi^2}m_b\; \ov{d}_i \sigma^{\mu\nu}(1+\gamma_5) T^a_{ij}G^a_{\mu\nu}
b_j,
\label{eq:o8g}
\eeq
where $i,j$ being the color indices of quarks. The corresponding effective Wilson
coefficient $C_{8g}^{eff}= C_{8g} + C_5$ \cite{nlo05}.

In Ref.~\cite{o8g2003}, the authors calculated the chromo-magnetic penguin contributions
to $B \to \phi K$ decays using the pQCD approach. They considered nine
chromo-magnetic penguin diagrams corresponding to the non-local operator
$O_{8g}^\prime$, as given in Eq.~(2.3) of Ref.~\cite{o8g2003}, generated by operator
$O_{8g}$ as defined in Eq.~(\ref{eq:o8g}).
The first two Feynman diagrams (a) and (b) in Ref.~\cite{o8g2003}
are the same as Figs.~(\ref{fig:mp})a and (\ref{fig:mp})b here.
According to Ref.~\cite{o8g2003}, the diagrams (a) and (b) dominate, while other seven
diagrams are small or negligible.
It is therefore reasonable for us to consider the NLO contributions induced by the
diagrams (a) and (b) only, for the sake of simplicity.

The decay amplitude for Figs.~\ref{fig:mp}a and \ref{fig:mp}b can be written  as
\beq
M^{(g)}_{K^*K}&=& \frac{4}{\sqrt{3}}G_F C_F^2m_B^6\int_{0}^{1}d
x_{1}d x_{2}\,d x_{3}\,\int_{0}^{\infty} b_1d b_1 b_2d
b_2\,\phi_B(x_1,b_1)\non
&&
\cdot \left\{ \left\{ -\left( 1-x_2 \right)
\left[ 2\pksa(\xeba)- \rks \left [ 3\pksp(\xeba)-\pkst(\xeba)\right]
\right. \right.\right.\non
&& \left. \left. \left.
-\rks x_2 \left[\pksp(\xeba)+\pkst(\xeba)\right] \right ]\pka(\xsba) \right.\right.\non
&& \left. \left.
+\rk \left( 1+x_2 \right)x_3 \cdot\pksa(\xeba)\left[ 3 \pkp(\xsba)+\pkt(\xsba)\right]
 \right.\right.\non
&& \left. \left.
-\rks\rk\left(1-x_2\right)\left[\pksp(\xeba)+\pkst(\xeba)\right]
\left [3\pkp(\xsba) -\pkt(\xsba)\right]
 \right.\right.\non
&& \left. \left.
-\rks\rk x_3\left(1-2x_2\right)\left[\pksp(\xeba)-\pkst(\xeba)\right]
\left[ 3\pkp(\xsba)+\pkt(\xsba)\right ] \right\}\right.\non &&
\left.
\hspace{1cm} \cdot E_g(t_q)h_g(A,B,C,b_1,b_2,b_3,x_2)
\right.\non &&  \left.
+ \left\{4\rks\pksp(\xeba)\pka(\xsba)
- 2\rks\rk x_3\pksp(\xeba)\left[ 3\pkp(\xsba)+\pkt(\xsba)\right] \right\}\right.\non
&& \left.
\hspace{1cm} \cdot E_g(t_q^{\prime})h_g(A^{\prime},B^{\prime},C^{\prime},b_2,b_1,b_3,x_1)
\right\},
\label{eq:mpp}
\eeq
for the case of $B \to K^*$ transition, and
\beq
M^{(g)}_{KK^*}&=&  \frac{4}{\sqrt{3}}G_F C_F^2m_B^6\int_{0}^{1}d
x_{1}d x_{2}\,d x_{3}\,\int_{0}^{\infty} b_1d b_1 b_2d
b_2\,\phi_B(x_1,b_1) \non
&&
\left\{ \left\{ -(1-x_2)\left[ 2\pka(\xeba)+ \rk \left [ 3\pkp(\xeba)-\pkt(\xeba)\right]
\right.\right.\right.\non
&& \left. \left.\left.
+\rk x_2 \left [\pkp(\xeba)+\pkt(\xeba)\right] \right ] \pksa(\xsba)\right.\right.\non
&& \left. \left.
+\rks\left(1+x_2\right)x_3 \pka(\xeba)\left[ 3 \pksp(\xsba)+\pkst(\xsba)\right]
\right.\right.\non&& \left. \left.
+\rk\rks\left(1-x_2\right)\left[ \pkp(\xeba)+\pkt(\xeba)\right]
\left [ 3\pksp(\xsba) -\pkst(\xsba)\right ]
\right.\right.\non
&& \left. \left.
+\rk\rks x_3\left(1-2x_2\right)\left[ \pkp(\xeba)-\pkt(\xeba)\right]
\left[ 3\pksp(\xsba)+\pkst(\xsba)\right] \right\} \right.\non &&\left.
 \hspace{1cm} \cdot E_g(t_q)h_g(A,B,C,b_1,b_2,b_3,x_2)
\right.\non && \left.
- \left\{4\rk\pkp(\xeba)\pksa(\xsba)-2\rk\rks x_3\pkp(\xeba)
\left[3\pksp(\xsba)+\pkst(\xsba)\right] \right\}
\right.\non && \left.
\hspace{1cm} \cdot
E_g(t_q^{\prime})h_g(A^{\prime},B^{\prime},C^{\prime},b_2,b_1,b_3,x_1)
\right\},
\label{eq:mpp2}
 \eeq
for the case of $B \to K$ transition. Here the hard scale $t_q$ and $t_q^\prime$ are
the same as in Eq.~(\ref{eq:tq-tqp}). The evolution factor $E_g(t)$ in Eqs.~(\ref{eq:mpp})
and (\ref{eq:mpp2}) is of the form
\beq
E_g(t)&=& C_{8g}^{eff}(t)\; \alpha_s^2(t)\cdot \exp\left [ -S_{g}\right ],
\eeq
with the Sudakov factor $S_{g}$ and the hard  function $h_g$,
\beq
S_{mg}(t) &=& s\left(x_1 m_B/\sqrt{2}, b_1\right)
 +s\left(x_2 m_B/\sqrt{2}, b_2\right)
+s\left((1-x_2) m_B/\sqrt{2}, b_2\right) \non
 && +s\left(x_3
m_B/\sqrt{2}, b_3\right) +s\left((1-x_3) m_B/\sqrt{2}, b_3\right)
\non
 &
&-\frac{1}{\beta_1}\left[\ln\frac{\ln(t/\Lambda)}{-\ln(b_1\Lambda)}
+\ln\frac{\ln(t/\Lambda)}{-\ln(b_2\Lambda)}+\ln\frac{\ln(t/\Lambda)}{-\ln(b_3\Lambda)}\right],
\label{eq:smg}
\eeq
\beq
h_g(A,B,C,&&b_1,b_2,b_3,x_i)=-S_t(x_i) \; K_0(Bb_1) \; K_0(Cb_3)\non
&& \cdot
\int_{0}^{\pi/2}d\theta \tan\theta\;
\cdot J_0(Ab_1\tan\theta)J_0(Ab_2\tan\theta)J_0(Ab_3\tan\theta), \ \
\eeq
where the functions $K_0(x)$ and $J_0(x)$ are the
Bessel functions, the form factor
$S_t(x_i)$ with $i=1,2$ has been given in Eq.~(\ref{eq:stxi}), and
the invariant masses $A^{(\prime)}, B^{(\prime)}$ and $C^{(\prime)}$ of the virtual
quarks and gluons are of the form
\beq
A &=&\sqrt{x_2}m_B,\quad B=B^{\prime}=\sqrt{x_1 x_2}m_B,\quad C=i\sqrt{(1-x_2)x_3}m_B,\non
A^{\prime}&=&\sqrt{x_1}m_B,\quad C^{\prime}=\sqrt{x_1-x_3}m_B.
\eeq

The total ``chromo-magnetic penguin" contribution to the considered $B \to KK^*$ decays
can therefore be written as
\beq
M_{KK^*}^{(cmp)} &=&  <K^* K|{\cal H}_{eff}^{cmp}|B>
= \lambda_t \; \left [ M^{(g)}_{K^*K} + M^{(g)}_{KK^*} \right],
\eeq
where $\lambda_t = V_{tb}V_{td}^{*}$.

From Fig.~(\ref{fig:mp}), one can see that the chromo-magnetic penguins
contribute only to $B^0\to K^{*0} \ov{K}^0 + K^{0} \ov{K}^{*0}$
and  $B^+\to K^{*+} \ov{K}^0, K^{+} \ov{K}^{*0}$ decays. For $B^0 \to K^{+} K^{*-}$ and
$K^{*+} K^{-}$ decays, there is again no such kind of NLO contributions.

\section{Numerical results and Discussions}\label{sec:n-d}

\subsection{Input parameters}

Besides those specified in the text, the following input parameters will also be used
in the numerical calculations:
\beq
m_B &=& 5.28 {\rm GeV}, \quad m_K=0.49{\rm GeV}, \quad m_{K^*}=0.892{\rm GeV},
\quad m_b=4.8 {\rm GeV},  \non
m_{0K}&=&1.7{\rm GeV}, \quad m_W = 80.41{\rm GeV}, \quad m_t = 168 {\rm GeV},
\quad \alpha_{em}=1/128, \non
f_B &=& 0.21 {\rm GeV}, \quad f_{K^*} = 0.217 {\rm GeV},\quad f_{K^*}^T = f_K =
0.16  {\rm GeV}, \non
\tau_{B^0} &=& 1.528 {\rm ps}, \quad \tau_{B^+} = 1.643 {\rm ps},
\label{para}
\eeq

For the CKM quark-mixing matrix, we use the Wolfenstein parametrization as given
in Ref.\cite{pdg2006}.
\beq
V_{ud}&=&0.9745, \quad V_{us}=\lambda = 0.2200, \quad |V_{ub}|=4.31\times 10^{-3},\non
V_{cd}&=&-0.224, \quad  V_{cd}=0.996, \quad V_{cb}=0.0413, \non
|V_{td}|&=& 7.4 \times 10^{-3}, \quad V_{ts}=-0.042, \quad |V_{tb}|=0.9991,
\label{eq:vckm}
\eeq
with the CKM angles $\beta=21.6^\circ$, $\gamma =60^\circ \pm 20^\circ $ and
$\alpha=100^\circ \pm 20^\circ $.
The unitarity condition $V_{ub} V_{uq}^* + V_{cb} V_{cq}^* + V_{tb} V_{tq}^* =0$ for $q=d,s$
is employed

\subsection{Branching ratios}

In the pQCD approach, the form factor $A_0^{B\to K^*}(q^2=0)$ and $F_{0,1}^{B\to K}(q^2=0)$
can be extracted from the decay amplitude $F_{eK^*}$ and $F_{eK^*}$ as shown in
Eqs.~(\ref{eq:ab-01}) and (\ref{eq:ab1}), via the following relations,
\beq
F_{0,1}^{B\to K}(q^2=0)&=&\frac{\sqrt{2}~F_{eK}} {G_F f_{K^*} m_B^2},  \label{eq:f01}\\
A_0^{B\to K^*}(q^2=0)&=&\frac{\sqrt{2}~F_{eK^*}} {G_F f_K m_B^2}.
\label{eq:a0bks}
\eeq
Consequently, one can find the NLO pQCD predictions for the
values of the corresponding form factors at zero momentum transfer:
 \beq
A_0^{B\to K^*}(q^2=0)&=& 0.38\pm0.05(\omega_b),
\quad
F^{B\to K}_{0,1}(q^2=0)=0.36 \pm 0.06(\omega_b),
\eeq
for $\omega_b=0.40\pm0.04$GeV, which agree well with those obtained in
QCD sum rule calculations, for example, in Refs.~\cite{ball98,ball05}.

For a general charmless two-body decays $B \to f$ with $f=M_2 M_3$, the branching ratio
can be written in general as
\beq
Br(B\to f) &=& \tau_B\; \frac{1}{16\pi m_B}\; \left | \calm \right|^2
\eeq
where $\tau_B$ is the lifetime of the B meson, and $ \calm= <K K^*| \calh_{eff}|B>$
for the case of $f= K\; K^*$.

Using  the wave functions and the input parameters as specified in
previous sections, it is straightforward  to calculate the branching ratios for the considered
decays. For $B^+ \to K^+ \ov K^{*0}$ and $B^+ \to K^{*+} \ov K^0$ decays, we show in
Table \ref{table1}, the CP-averaged branching ratios
\beq
Br(B \to f ) = \frac{1}{2}\left [ Br(B \to f ) + Br(\ov{B}\to \ov{f} )\right].
\eeq

For $B^0$ decays, it is a little complicate since both $B^0$ and $\ov B^0$ can
decay into the final state $f$ and $\ov f$ simultaneously.
In Table \ref{table1}, we show the CP-averaged Br's for $B^0 \to f_1$, $B^0 \to \bar{f}_1$ and
for $B^0 \to f_1 + \bar{f}_1$ with $f_1=K^0 \ov{K}^{*0}$, respectively.
The third result corresponds to the measured upper
limit. For $B^0 \to f_2, \bar{f}_2$ and $B^0 \to f_2 + \bar{f}_2$ with
$f_2= K^+ K^{*-} $, we take the same convention.

Except for the LO results, we always use the NLO Wilson coefficients in the
calculations. The label $+$VC, $+$QL, $+$MP and NLO denote the pQCD predictions with
the inclusion of the vertex corrections only, the quark loops only,
the magnetic-penguin only, and all the considered NLO corrections, respectively.
For the sake of comparison, we also show currently available experimental results
\cite{hfag} and the numerical results evaluated in
the framework of the QCD factorization (QCDF) \cite{npb675}.

\begin{table}[thb]
\begin{center}
\caption{ The pQCD predictions for the branching ratios
(in unit of $10^{-7}$).
The label $\rm{LO}$ means the leading-order results, and $+$VC, $+$QL, $+$MP, NLO
mean the predictions with the inclusion of the vertex corrections, the quark loops,
the magnetic-penguin, and all the considered NLO corrections, respectively.
 }
\label{table1}
\vspace{0.2cm}
\begin{tabular}{l  |c c c c c |c |c } \hline \hline
 Mode&  LO & +VC & +QL & +MP  & NLO & Data & QCDF\\
\hline
$B^+ \to K^+ \overline{K}^{*0}$ &4.2&5.3&5.8&3.1&3.2& $ < 11 $& $3.0^{+6.0}_{-2.5}$ \\
$B^+ \to {K}^{*+}\overline{K}^0$ &2.0&2.7&2.3&1.6&2.1& &$3.0^{+7.2}_{-2.7}$ \\ \hline
$B^0 \to K^0 \ov{K}^{*0}$ &2.1 &3.0 &2.9 &1.8&2.4 &$-$ & $2.6^{+2.8}_{-2.0}$ \\
$B^0 \to \ov K^0 K^{*0} $ &6.4&6.9&8.0 &4.3& 4.9&$-$ &$2.9^{+7.3}_{-2.7}$ \\
$B^0 \to K^0 \ov{K}^{*0} + K^0 K^{*0} $ &13.7&  14.0 & 15.2& 6.7 &8.5&$ < 19 $& \\ \hline
$B^0 \to K^+ \ov{K}^{*-}$ &1.1 &$-$&$-$&$-$&0.83&&$0.14^{+1.07}_{-0.14}$ \\
$B^0 \to K^- \ov K^{*+}$  &0.41& $-$&$-$&$-$&0.17&&$0.14^{+1.07}_{-0.14}$  \\
$B^0 \to K^+ \ov{K}^{*-} + K^- \ov K^{*+}$ &2.7 & $-$&$-$&$-$&1.3&&\\
\hline \hline
\end{tabular}
\end{center} \end{table}

It is worth stressing that the theoretical predictions in the pQCD
approach have relatively large theoretical errors induced by the
still large uncertainties of many input parameters.
The pQCD predictions for the branching ratios with the consideration of major
uncertainties are the following (in unit of $10^{-7}$)
\beq
Br(\ B^+ \to K^+ \ov{K}^{*0}) &=& 3.2
^{+1.0}_{-0.6}(\omega_b)^{+0.2}_{-0.1}(\alpha)^{+0.5}_{-0.3}(a_{iK})
^{+0.2}_{-0.1}(a_{iK^*}) ,\non
Br(\ B^+ \to {K}^{*+}\ov{K}^0) &=& 2.1
^{+0.1}_{-0.2}(\omega_b)^{+0.2}_{-0.3}(\alpha)^{+1.3}_{-1.2}(a_{iK})
\pm0.4(a_{iK^*}), \non
Br(\ B^0 \to K^0\overline{K}^{*0}) &=& 2.4
\pm0.2(\omega_b)^{+0.0}_{-0.1}(\alpha)^{+0.3}_{-0.4}(a_{iK})
^{+0.6}_{-0.4}(a_{iK^*}) \label{eq:brp-etap},\non
Br(\ B^0 \to \ov {K}^0 K^{*0}) &=& 4.9
^{+1.2}_{-0.8}(\omega_b)^{+0.3}_{-0.2}(\alpha)^{+0.5}_{-0.4}(a_{iK})
^{+0.3}_{-0.1}(a_{iK^*}) \label{eq:brp-eta},\non
Br(B^0 \to f_1 + \bar{f}_1) &=& 8.5
^{+2.2}_{-1.7}(\omega_b)\pm0.1(\alpha)^{+1.0}_{-0.9}(a_{iK})
^{+1.1}_{-0.6}(a_{iK^*}) \label{eq:brp-etab}, \non
Br(\ B^0 \to K^+ K^{*-}) &=& 0.83 ^{+0.04}_{-0.08}(\omega_b)\pm0.6(\alpha)^{+0.28}_{-0.21}(a_{iK})
^{+0.28}_{-0.18}(a_{iK^*})\label{eq:br0-eta} ,\non
Br(\ B^0 \to K^- K^{*+}) &=&0.17^{+0.02}_{-0.01}(\omega_b)\pm0.07(\alpha)^{+0.26}_{-0.08}(a_{iK})
\pm0.02(a_{iK^*}) \label{eq:br0-etap},\non
Br(B^0 \to f_2 + \bar{f}_2) &=& 1.3
\pm 0.1(\omega_b)^{+0.0}_{-0.1}(\alpha)^{+0.2}_{-0.6}(a_{iK})
^{+0.4}_{-0.3}(a_{iK^*}).\label{eq:br-kk2}
\eeq
The major theoretical errors are induced by the uncertainties of
$\omega_b=0.4 \pm 0.04$ GeV, $\alpha = 100^\circ \pm 20^\circ$, and Gegenbauer
coefficients $a_{1K}=0.17\pm0.17$, $a_{2K}=0.115\pm0.115$; $a_{1K^*}=0.03\pm0.03$,
$a_{2K^*}=0.11\pm0.11$, respectively.
Additionally, the final-state interactions remains unsettled in
pQCD, which is non-perturbative but not universal. Fortunately, good
agreement between the pQCD predictions for the branching ratios of
$B\to K K$ decays \cite{ch2000} and currently available
experimental measurements \cite{hfag} indicates that the FSI effects
are most possibly not important.

From the numerical results, it is easy to see that
\begin{itemize}

\item
The pQCD predictions for the branching ratios of $B^+\to K^+ \overline{K}^{*0}$
and $B^0 \to K^0 \ov{K}^{*0} + K^0 K^{*0}$ are consistent with currently available
experimental upper limits. Inclusion of the NLO contributions decreases the central value of the
LO predictions by about $30\%$ to $80\%$. The chromo-magnetic
penguin provide the dominant NLO contributions.

\item
For $B^0 \to K^+ K^{*-}$ decay, the pQCD prediction is rather different from that from
the  QCD factorization approach. Such difference could be tested in the forthcoming LHCb experiments.
For other decays, the pQCD predictions agree well with the corresponding QCDF results within one
standard deviation.

\end{itemize}

\subsection{CP-violating asymmetries}

   Now we turn to the evaluations of the CP-violating asymmetries of
$B \to K^*K$ decays in pQCD approach. For $B^+ \to K^+\overline{K}^{*0}$
and $B^+ \to K^{*+}\overline{K}^0$ decays, the
direct CP-violating asymmetries $\acp$ can be defined as:
\beq
\acp^{dir} =  \frac{|\overline{\cal M}_f|^2 - |{\cal M}_f|^2}{
 |\overline{\cal M}_f|^2+|{\cal M}_f|^2}, \label{eq:acp1}
 \eeq
where $\calm_f =<f|\calh_{eff}|B>$ and $\ov\calm_f =<\bar{f}|\calh_{eff}|B>$.

The pQCD predictions for the direct CP-violating asymmetries of the considered decays are
listed in Table \ref{table-acp1}.
For comparison, we also reproduce verbatim the corresponding
numerical results evaluated in the framework of the QCD
factorization (QCDF) \cite{npb675}.

\begin{table}[thb]
\begin{center}
\caption{ The pQCD predictions for the direct CP asymmetries of $B \to KK^*$ decays
(in units of percent). }
\label{table-acp1}
\vspace{0.2cm}
\begin{tabular}{l  |c c c c c |c } \hline \hline
 Mode  &  LO &  +VC & +QL & +MP &  NLO &\hspace{0.5cm}QCDF\hspace{0.5cm}\\
\hline
$\acp^{dir}(B^+ \to K^+\overline{K}^{*0})$ &-53.4 &-39.2&-5.1&-49.8&-6.9&$-24^{+28}_{-39}$\\ \hline
$\acp^{dir}(B^+ \to K^{*+}\overline{K}^0)$ &8.1 &-12.3&7.2&10.6&6.5&$-13^{+29}_{-37}$\\
 \hline \hline
\end{tabular}
\end{center} \end{table}

The pQCD predictions for $\acp^{dir}$ and the major theoretical
errors for $B^+ \to K^+\overline{K}^{*0}, K^{*+}\overline{K}^0$
decays are
\beq
\acp^{dir}(B^+ \to K^+\overline{K}^{*0}) &=& \left
[-6.9^{+5.6}_{-5.3}(\omega_b)^{+1.0}_{-0.3}(\alpha)
^{+9.2}_{-6.5}(a_{iK})^{+4.0}_{-6.0}(a_{iK^*})\right ]
 \times 10^{-2} \label{eq:acp-a},\quad \\
\acp^{dir}(B^+ \to K^{*+}\overline{K}^0) &=& \left [
6.5^{+7.9}_{-7.3}(\omega_b)^{+1.1}_{-1.4}(\alpha)
^{+9.1}_{-7.7}(a_{iK})^{+2.1}_{-3.9}(a_{iK^*}) \right ] \times
10^{-2} \label{eq:acp-b},
\eeq
where the dominant errors come from
the variations of $\omega_b=0.4 \pm 0.04$ GeV, $\alpha = 100^\circ
\pm 20^\circ$, and Gegenbauer coefficients $a_{1K}=0.17\pm0.17$,
$a_{2K}=0.115\pm0.115$; $a_{1K^*}=0.03\pm0.03$,
$a_{2K^*}=0.11\pm0.11$, respectively.

\begin{figure}[t,b]
\begin{center}
\includegraphics[scale=0.65]{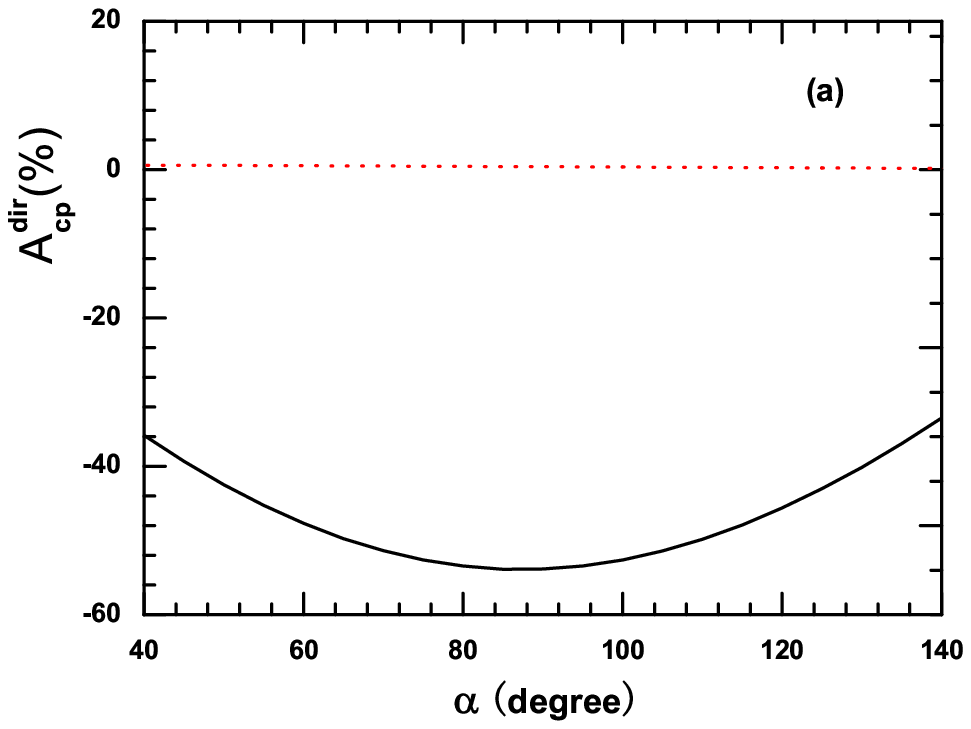}
\includegraphics[scale=0.65]{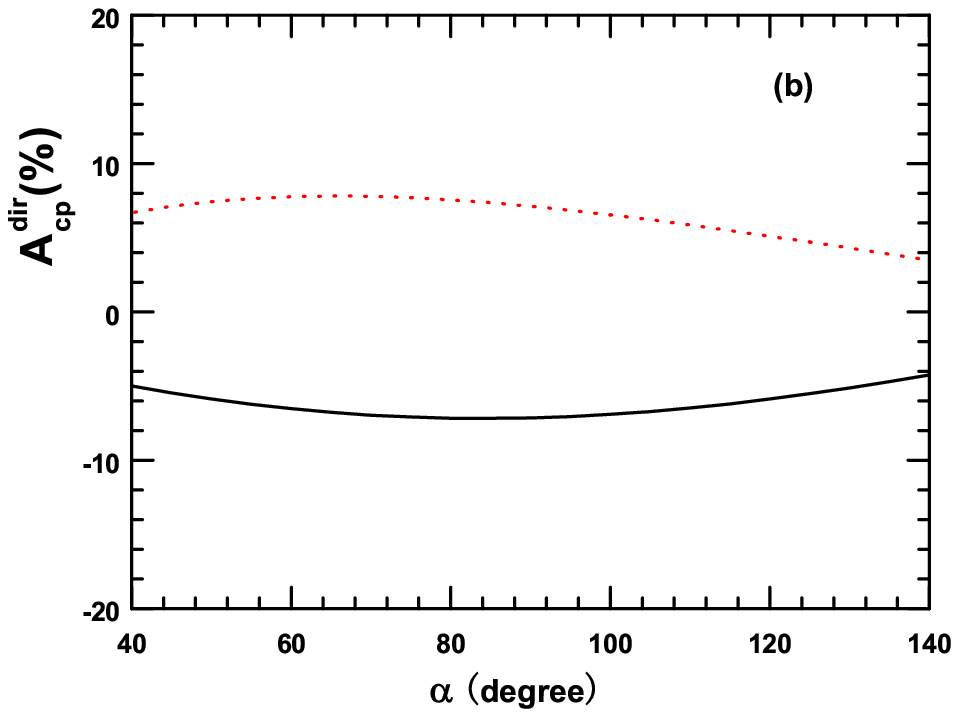}
\vspace{0.3cm} \caption{The direct CP asymmetries (in percentage) of
$B^+\to K^+\overline{K}^{*0}$ (solid curve) and $B^+\to
{K}^{*+}\overline{K}^0$ (dotted curve) as a function of CKM angle
$\alpha$. (a) shows the LO results, while (b) shows the NLO results. }
\label{fig:kzdir}
\end{center}
\end{figure}

In Fig.~\ref{fig:kzdir}, we show the $\alpha-$dependence
of the direct CP-violating asymmetries $\acp^{dir}$ for $B^+ \to
K^+\overline{K}^{*0}$ (the solid curve) and $B^+ \to
K^{*+}\overline{K}^0$ (the dotted curve) decay, respectively. The
left figure is for the LO pQCD predictions and the right one for the
NLO pQCD predictions. One can see from the numbers and figures that
(a) as usual, there exist relatively large differences between the pQCD and QCDF
predictions;
(b) the LO and NLO pQCD predictions for the direct CP-violating asymmetries
are also rather different; and (c) the NLO contribution from the "Quark-loops"
("Vertex corrections ") leads to the dominate change of $\acp^{dir}$
for $B^+ \to K^{+}\overline{K}^{*0}$ ( $B^+ \to K^{*+}\overline{K}^0$) decay.

We now study the CP-violating asymmetries for $B^0/\ov B^0 \to K^+
K^{*-}(K^- K^{*+})$ decays. Since both $B^0$ and $\ov B^0$ can
decay into the final state $K^+K^{*-}$ and $K^- K^{*+}$, the
four time-dependent
decay widths for $B^0(t) \to K^+ K^{*-}$, $\ov B^0(t) \to K^-
K^{*+}$, $B^0(t) \to K^- K^{*+}$ and $\ov B^0(t) \to K^+ K^{*-}$
can be expressed by four basic matrix elements:
\beq
 g= \langle K^+ K^{*-} |H_{eff}| B^0\rangle ,\; \;\; \;h=\langle K^+ K^{*-} |H_{eff}| \overline
 B^0 \rangle,
\non \overline g= \langle K^- K^{*+} |H_{eff}| \overline B^0\rangle
,\;\;\; \; \overline h=\langle K^- K^{*+} |H_{eff}|
 B^0\rangle,
\eeq
which determine the decay matrix elements of $B^0 \to K^+ K^{*-}$, $\ov B^0
\to K^- K^{*+}$, $B^0 \to K^- K^{*+}$ and $\ov B^0 \to K^+ K^{*-}$ at $t=0$.
Besides the matrix elements $g, \bar{g}, h$ and $\bar{h}$, one also need to know the
CP-violating parameter coming from the $B^0-\ov B^0$ mixing:
\beq
 B_1=p|B^0\rangle+q| \overline B^0\rangle,\;\; \; \;B_2=p|B^0\rangle-q| \overline
 B^0\rangle,
\eeq
with $|p|^2 + |q|^2 = 1$.

Following the notation of Ref. \cite{guo07}, the four time-dependent
widths are given by the following formulae:
 \beq
 \Gamma(B^0 (t)\to K^+ K^{*-})&=&e^{-\Gamma t}\frac
 {1}{2}(|g|^2+|h|^2)\times \left \{1+ a_{\epsilon'} \cos(\Delta mt)
 +a_{\epsilon+\epsilon'}\sin(\Delta mt)\right \},
\non \Gamma(\ov B^0 (t)\to K^- K^{*+})&=&e^{-\Gamma t}\frac
 {1}{2}(|\ov g|^2+|\ov h|^2)\times \left \{1- a_{\ov \epsilon'} \cos(\Delta mt)
 -a_{\epsilon+\ov \epsilon'}\sin(\Delta mt)\right \},
 \non \Gamma(B^0 (t)\to K^- K^{*+})&=&e^{-\Gamma t}\frac
 {1}{2}(|\ov g|^2+|\ov h|^2)\times \left \{1- a_{\ov \epsilon'} \cos(\Delta mt)
 -a_{\epsilon+\ov \epsilon'}\sin(\Delta mt)\right \},
 \non \Gamma(\ov B^0 (t)\to K^+ K^{*-})&=&e^{-\Gamma t}\frac
 {1}{2}(|g|^2+|h|^2)\times \left \{1+ a_{\epsilon'} \cos(\Delta mt)
 +a_{\epsilon+\epsilon'}\sin(\Delta mt)\right \},
 \eeq
 where the CP -violating parameters are
 \beq
 a_{\epsilon'}&=&\frac{|g|^2-|h|^2}{|g|^2+|h|^2},\quad
 a_{\epsilon+\epsilon'}=\frac{-2Im(\frac{q}{p}\frac{h}{g})}{1+|h/g|^2} \non
 a_{\ov \epsilon'}&=&\frac{|\ov h|^2-|\ov g|^2}{|\ov h|^2+|\ov g|^2},\quad
 a_{\epsilon+\ov \epsilon'}=\frac{-2Im(\frac{q}{p}\frac{\ov g}{\ov h})}{1+|\ov g/\ov
 h|^2}, \label{eq:a41}
 \eeq
with $q/p=e^{-2i\beta}$a and $\beta=21.6^\circ$ is one of the three CKM angles.

\begin{figure}[t,b]
\begin{center}
\includegraphics[scale=0.7]{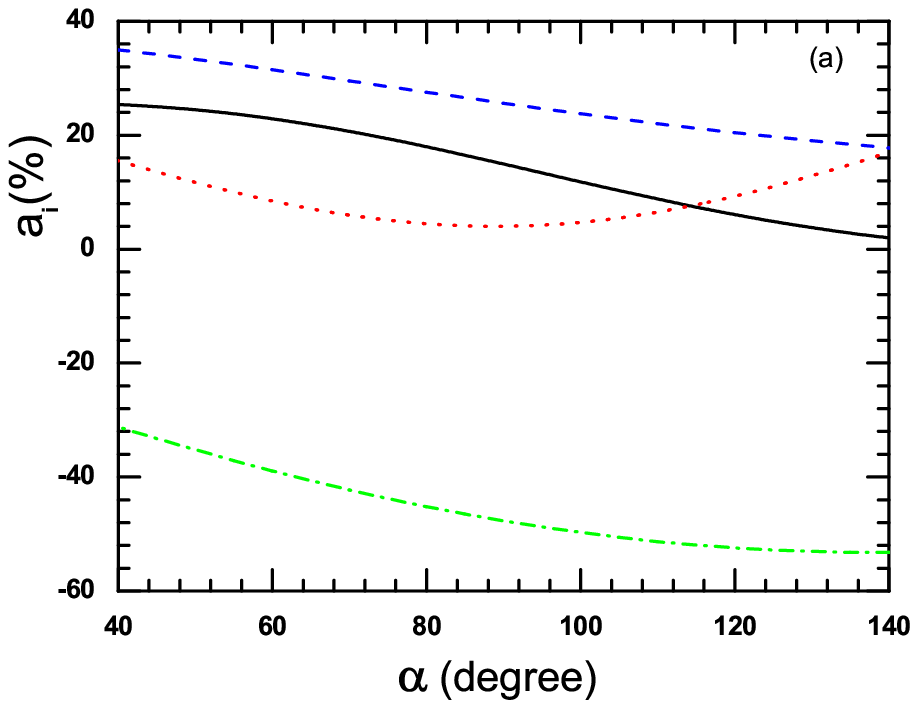}
\includegraphics[scale=0.7]{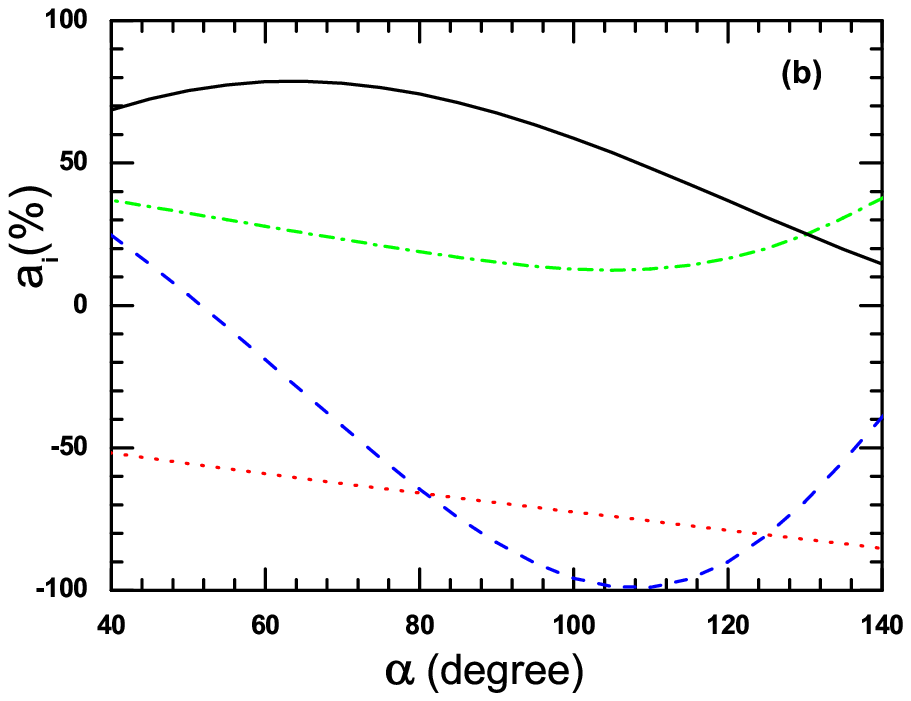}
\vspace{0.3cm}
\caption{The CP-violating parameters of $B^0/\ov B^0
\to K^0\overline{K}^{*0} (K^0 K^{*0})$ decays,(a);  and
$B^0/ \ov B^0 \to K^+K^{*-}(K^-K^{*+})$ decays, (b) :
$a_{\epsilon'}$ (dash-dotted line), $a_{\ov \epsilon'}$ (dotted line),
$a_{\epsilon+\epsilon'}$ (dashed line) and $a_{\epsilon+\ov
\epsilon'}$ (solid line) as a function of CKM angle $\alpha$.
}\label{fig:aepxi}
\end{center}
\end{figure}

Similarly, the four time-dependent decay widths for $B^0
\to K^0 \overline{K}^{*0}$, $\ov B^0 \to K^0 \overline{K}^{*0}$,
$B^0 \to \ov K^0  K^{*0}$ and $\ov B^0 \to \ov K^0 K^{*0}$ (here $K^0$ means $K^0_S$)
can also be defined as
\beq
 g &=& \langle K^0 \overline{K}^{*0} |H_{eff}| B^0\rangle ,\quad
 h=\langle K^0 \overline{K}^{*0} |H_{eff}| \overline  B^0 \rangle,\non
 \ov g &=& \langle \ov K^0 K^{*0} |H_{eff}| \overline B^0\rangle ,\quad
 \overline h=\langle \ov K^0 K^{*0} |H_{eff}| B^0\rangle,
\eeq
One can define, consequently, the four CP-violating parameters $a_{\epsilon'}$,
$a_{\epsilon+\epsilon'}$, $a_{\ov \epsilon'}$ and $ a_{\epsilon+\ov \epsilon'}$ for
$B^0/\bar{B}^0 \to f_1 + \bar{f}_1$ decays in the same way as in Eq.~(\ref{eq:a41}).
In Fig.~\ref{fig:aepxi}, we show the pQCD predictions for
the eight CP-violating parameters for the considered decays.

The central values of the pQCD predictions for the CP-violating parameters are
\beq
a_{\epsilon'}=0.13, \quad a_{\epsilon+\epsilon'}=-0.96,\quad
a_{\ov \epsilon'}=-0.72, \quad a_{\epsilon+\ov \epsilon'}=0.59.
\eeq
for $B^0/\ov{B}^0 \to K^+ K^{*-} + K^{-} K^{*+}$ decays, and
\beq
a_{\epsilon'}=-0.50, \quad a_{\epsilon+\epsilon'}=0.24,\quad
a_{\ov \epsilon'}=0.05, \quad a_{\epsilon+\ov \epsilon'}=0.12,
\eeq
for $B^0/\ov{B}^0 \to K^0 \ov K^{*0} + \ov K^{0} K^{*0}$ decays.

As pointed in Ref.~\cite{beneke07}, it may be conceptually incorrect to evaluate
the Wilson coefficients at scales down to $0.5$ GeV. The explicit numerical values
for the Wilson coefficients $C_1(\mu) - C_{10}(\mu)$ for $\mu=0.5, 1.0, 1.5$ and $2.0$
GeV, as listed in Table \ref{wc-mu}, also support this expectation: the values of the
Wilson coefficients $C_{3,4,5,6}(\mu)$ at $\mu=0.5$ GeV are about four to seven times
larger than those at $\mu=1.0$ GeV. For $C_5(\mu)$, specifically, $C_5(0.5)$ and $C_5(1.0)$
even have a different sign besides the large difference in their magnitude.
In the region of $\mu \geq 1.0$ GeV, however,
the $\mu-$dependence of all Wilson coefficients become relatively weak.
It is therefore reasonable for us to choose $\mu_0=1.0$ GeV as the lower cut-off of the hard
scale, instead of $\mu_0=0.5$ GeV as being assumed in Ref.~\cite{nlo05}.
We then fix the values $C_{i}(\mu)$ at $C_{i}(\mu_0=1.0)$, whenever the scale $\mu$ runs to
below the scale $\mu_0$.

In order to show directly the $\mu_0$-dependence of the branching ratios and CP-violating asymmetries,
we recalculated these quantities for $B \to KK^*$ decays by setting $\mu_0=0.5$, $1.0, 1.5$
and $2.0$ GeV, respectively.
It is easy to see  from the numerical results as listed in Table \ref{br-muc}  that
the pQCD predictions are relatively stable against the variation of $\mu_0$
for $\mu_0 \geq 1.0$ GeV. We therefore set $\mu_0=1.0$ GeV to be the cut-off scale for Wilson
coefficients $C_i(\mu)$.
Of course, the issue of $\mu_0$-dependence need more studies.

\begin{table}[thb]
\begin{center}
\caption{NLO Wilson coefficients $C_i(\mu)$ for $\mu_0=0.5-2.0$ GeV,
respectively.}
\label{wc-mu} \vspace{0.2cm}
\begin{tabular}{l|ll|rrrr|rrrr } \hline \hline
$\mu_0$ & $C_1(\mu)$ & $C_2(\mu)$ & $C_3(\mu)$ & $C_4(\mu)$ & $C_5(\mu)$ & $C_6(\mu)$&
$C_7(\mu)$&$C_8(\mu)$&$C_9(\mu)$&$C_{10}(\mu)$   \\ \hline
0.5 GeV &$-0.9923$&$1.6537$&$0.1729$&$-0.3122$&$-0.1143$&$-0.8276$&$ 0.0010$&$0.0056$&$-0.0148$&$0.0092$ \\ \hline
1.0 GeV &$-0.5093$&$1.2790$&$0.0428$&$-0.0898$&$ 0.0150$&$-0.1321$&$-0.0002$&$0.0008$&$-0.0120$&$0.0050$ \\ \hline
1.5 GeV &$-0.3773$&$1.1920$&$0.0289$&$-0.0652$&$ 0.0153$&$-0.0856$&$-0.0002$&$0.0005$&$-0.0112$&$0.0038$ \\ \hline
2.0 GeV &$-0.3114$&$1.1518$&$0.0230$&$-0.0541$&$ 0.0145$&$-0.0672$&$-0.0002$&$0.0004$&$-0.0108$&$0.0032$ \\ \hline
\hline
\end{tabular} \end{center} \end{table}

\begin{table}[thb]
\begin{center}
\caption{The pQCD predictions for the branching ratios (in unit of $10^{-7}$) and
direct CP-violating asymmetries (in unit of $10^{-2}$) for the considered
$B \to K K^*$ decays, assuming $\mu_0=0.5,1.0,1.5$ and $2.0$ GeV, respectively.}
\label{br-muc} \vspace{0.2cm}
\begin{tabular}{l|c|c|c|c } \hline \hline
 Mode&  $\mu_0=0.5$ & $\mu_0=1.0$ & $\mu_0=1.5$ & $\mu_0=2.0$ \\
\hline
$Br(B^+ \to K^+ \overline{K}^{*0})$     &4.7 &3.2&2.6 & 2.1 \\
$Br(B^+ \to K^{*+}\overline{K}^0)$     &2.6 &2.1&1.3 &0.8 \\
$Br(B^0/\ov{B}^0 \to f_1 + \bar{f}_1)$ &22.5&8.5&5.0 &3.5 \\
$Br(B^0/\ov{B}^0 \to f_2 + \bar{f}_2)$ &5.4 &1.3&0.78 &0.55\\ \hline
$\acp^{dir}(B^+ \to K^+\overline{K}^{*0})$ &-4.0&-6.9&-7.1&-5.1 \\
$\acp^{dir}(B^+ \to K^{*+}\overline{K}^0)$ &16.8&6.5 &-1.5 &-5.8 \\
\hline \hline
\end{tabular} \end{center} \end{table}

\section{summary }

In this paper, we calculate some NLO contributions to the branching ratios and
CP-violating asymmetries of $B \to K K^*$ decays   in the
pQCD factorization approach.

From our calculations and phenomenological analysis, we found the following results:
\begin{itemize}
\item
The NLO contributions from the QCD vertex corrections, the quark-loops and the chromo-magnetic
penguins can be rather large and provide  significant
modifications to the LO predictions.

\item
The NLO pQCD predictions for the form factors of $B\to K^*$ and $K$ transitions are
\beq
A_0^{B\to K^*}(q^2=0)&=& 0.38\pm0.05(\omega_b),\non
F^{B\to K}_{0,1}(q^2=0)&=&0.36 \pm 0.06 (\omega_b),
\eeq
for $\omega_b=0.40\pm0.04$GeV, which agree well with those obtained in
QCD sum rule calculations.

\item
The pQCD predictions for the branching ratios are
\beq
Br( B^+ \to \overline{K}^{*0}K^+) &=&3.2 ^{+1.2}_{-0.8} \times 10^{-7} ,\non
Br( B^+ \to {K}^{*+}\overline{K}^0) &=& 2.1 ^{+1.4}_{-1.2}\times 10^{-7}, \non
Br(B^0 \to K^0\overline{K}^{*0}+\overline{K}^0 K^{*0}) &=&
8.5 ^{+2.6}_{-2.1} \times 10^{-7},\non
Br( B^0 \to K^+K^{*-}+K^-K^{*+}) &=&1.3 ^{+0.5}_{-0.7} \times 10^{-7},
\eeq
where the theoretical errors from various sources are added in quadrature.
These pQCD predictions are consistent  with both the QCDF predictions and
currently available experimental upper limits.

\item
The direct CP-violating asymmetries for $B^+ \to K^+\overline{K}^{*0},
K^{*+}\overline{K}^0$ are (in unit of $10^{-2}$)
\beq
\acp^{dir}(B^+ \to K^+\overline{K}^{*0}) &=& -6.9^{+11.5}_{-10.3},\non
\acp^{dir}(B^+ \to K^{*+}\overline{K}^0) &=& 6.5^{+12.3}_{-11.4},
\eeq
which are rather different from those in the QCDF approach.

\end{itemize}

\begin{acknowledgments}

The authors are very grateful  to Hsiang-nan Li, Cai-Dian L\"u, Ying Li, Wei Wang
and Yu-Ming Wang for helpful discussions.
This work is partly supported  by the National Natural Science
Foundation of China under Grant No.10575052 and 10735080.
\end{acknowledgments}

\begin{appendix}

\section{Related Functions }\label{sec:aa}

We show here the function $h_i$'s, coming from the Fourier
transformations  of $H^{(0)}$,
\beq
h_e(x_1,x_2,b_1,b_2)&=&  K_{0}\left(\sqrt{x_1 x_2} m_B b_1\right)
\left[\theta(b_1-b_2)K_0\left(\sqrt{x_2} m_B
b_1\right)I_0\left(\sqrt{x_2} m_B b_2\right)\right. \non & &\;\left.
+\theta(b_2-b_1)K_0\left(\sqrt{x_2}  m_B b_2\right)
I_0\left(\sqrt{x_2}  m_B b_1\right)\right] S_t(x_2),
\label{eq:he}
\eeq
\beq
h_a(x_2,x_3,b_2,b_3)&=& K_{0}\left(i\sqrt{(1-x_2)x_3} m_B b_2\right)
 \left[\theta(b_3-b_2)K_0\left(i \sqrt{x_3} m_B
b_3\right)I_0\left(i \sqrt{x_3} m_B b_2\right)\right. \non
& &\;\;\;\;\left. +\theta(b_2-b_3)K_0\left(i \sqrt{x_3}  m_B
b_2\right) I_0\left(i \sqrt{x_3}  m_B b_3\right)\right] S_t(x_3),
\label{eq:ha}
\eeq
\beq
h_{f}(x_1,x_2,x_3,b_1,b_3) &=& \biggl\{\theta(b_1-b_3)\mathrm{K}_0(m_B\sqrt{x_1 x_2} b_1)
  \mathrm{I}_0(m_B\sqrt{x_1 x_2} b_3)\non &+ & \theta(b_3-b_1)\mathrm{K}_0(m_B\sqrt{x_1 x_2} b_3)
  \mathrm{I}_0(m_B\sqrt{x_1 x_2} b_1) \biggr\} \non && \cdot\left(
\begin{matrix}
 \frac{\pi i}{2}\mathrm{H}_0(\sqrt{(x_2(x_3-x_1))} m_B b_3), & \text{for}\quad x_1-x_3<0 \\
 \mathrm{K}_0^{(1)}(\sqrt{(x_2(x_1-x_3)}m_B b_3), &
 \text{for} \quad x_1-x_3>0
\end{matrix}\right),
\label{eq:hf}
\eeq
\beq
h_f^3(x_1,x_2,x_3,b_1,b_3) &=&
\biggl\{\theta(b_1-b_3) \mathrm{K}_0(i \sqrt{(1-x_2) x_3} b_1 M_m)
 \mathrm{I}_0(i \sqrt{(1-x_2) x_3} b_3 m_B)\non &+&(\theta(b_3-b_1) \mathrm{K}_0(i \sqrt{(1-x_2) x_3} b_3 m_B)
 \mathrm{I}_0(i \sqrt{(1-x_2) x_3} b_1 m_B) \biggr\}
 \non
& & \cdot\left(
\begin{matrix}
 \mathrm{K}_0(m_B\sqrt{(x_1-x_3)(1-x_2)}b_1), & \text{for}\quad x_1-x_3>0 \\
 \frac{\pi i}{2} \mathrm{H}_0^{(1)}(m_B\sqrt{(x_3-x_1)(1-x_2)}b_1), &
 \text{for} \quad x_1-x_3<0
\end{matrix} \right),
\label{eq:hf3}
\eeq
 \beq
 h_f^4(x_1,x_2,x_3,b_1,b_2) &=&
 \biggl\{\theta(b_1-b_3) \mathrm{K}_0(i \sqrt{(1-x_2) x_3} b_1 m_B)
 \mathrm{I}_0(i \sqrt{(1-x_2) x_3} b_3 m_B)
 \non
&+& \theta(b_3-b_1) \mathrm{K}_0(i \sqrt{(1-x_2) x_3} b_3
M_B)\mathrm{I}_0(i \sqrt{(1-x_2) x_3} b_1 m_B)\biggr\} \non &&\cdot
\left(
\begin{matrix}
 \mathrm{K}_0(m_B F_{1}b_1), & \text{for}\quad F_{1}^2>0 \\
 \frac{\pi i}{2} \mathrm{H}_0^{(1)}(m_B \sqrt{|F_{1}^2|}b_1), &
 \text{for}\quad F_{1}^2<0
\end{matrix}\right),
\label{eq:hf4}
\eeq
where $J_0$ is the Bessel function and $K_0$, $I_0$ are
modified Bessel functions $K_0 (-i x) = -(\pi/2) Y_0 (x) + i (\pi/2)
J_0 (x)$, and $F_{(1)}$'s are defined by
\beq
F^2_{(1)}&=&1-x_2(1-x_3-x_1).
\eeq

The threshold resummation form
factor $S_t(x_i)$ is adopted from Ref.\cite{kurimoto}. It has been
parametrized as \beq S_t(x)=\frac{2^{1+2c} \Gamma
(3/2+c)}{\sqrt{\pi} \Gamma(1+c)}[x(1-x)]^c,
\label{eq:stxi}
\eeq
where the parameter $c=0.3$.

The evolution factors $E^{(\prime)}_e$ and $E^{(\prime)}_a$ appeared in
Eqs.~(\ref{eq:ab-01}) to (\ref{eq:cd-05}) are given by
\beq
E_e(t)&=&\alpha_s(t)\exp[-S_{ab}(t)],\non
E_e^{\prime}(t)&=&\alpha_s(t)\exp[-S_{cd}(t)]|_{b_2=b_1},\non
E_a(t)&=&\alpha_s(t)\exp[-S_{gh}(t)],\non
E_a^{\prime}(t)&=&\alpha_s(t)\exp[-S_{ef}(t)]|_{b_2=b_3},
 \eeq
where the Sudakov factors can be written as
\beq
S_{ab}(t) &=& s\left(x_1 m_B/\sqrt{2}, b_1\right) +s\left(x_2 m_B/\sqrt{2},
b_2\right) +s\left((1-x_2) m_B/\sqrt{2}, b_2\right) \non
&&-\frac{1}{\beta_1}\left[\ln\frac{\ln(t/\Lambda)}{-\ln(b_1\Lambda)}
+\ln\frac{\ln(t/\Lambda)}{-\ln(b_2\Lambda)}\right],
\label{eq:sab}\\
 S_{cd}(t) &=& s\left(x_1 m_B/\sqrt{2}, b_1\right)
 +s\left(x_2 m_B/\sqrt{2}, b_1\right)
+s\left((1-x_2) m_B/\sqrt{2}, b_1\right) \non
 && +s\left(x_3
m_B/\sqrt{2}, b_3\right) +s\left((1-x_3) m_B/\sqrt{2}, b_3\right)
\non
 & &-\frac{1}{\beta_1}\left[2
\ln\frac{\ln(t/\Lambda)}{-\ln(b_1\Lambda)}
+\ln\frac{\ln(t/\Lambda)}{-\ln(b_3\Lambda)}\right],
\label{eq:scd}\\
S_{ef}(t) &=& s\left(x_1 m_B/\sqrt{2}, b_1\right)
 +s\left(x_2 m_B/\sqrt{2}, b_2\right)
+s\left((1-x_2) m_B/\sqrt{2}, b_2\right) \non
 && +s\left(x_3
m_B/\sqrt{2}, b_2\right) +s\left((1-x_3) m_B/\sqrt{2}, b_2\right)
\non
&&
-\frac{1}{\beta_1}\left[\ln\frac{\ln(t/\Lambda)}{-\ln(b_1\Lambda)}
+2\ln\frac{\ln(t/\Lambda)}{-\ln(b_2\Lambda)}\right],
\label{eq:sef}\\
S_{gh}(t) &=& s\left(x_2 m_B/\sqrt{2}, b_2\right)
 +s\left(x_3 m_B/\sqrt{2}, b_3\right)
+s\left((1-x_2) m_B/\sqrt{2}, b_2\right) \non
 &+& s\left((1-x_3)
m_B/\sqrt{2}, b_3\right)
-\frac{1}{\beta_1}\left[\ln\frac{\ln(t/\Lambda)}{-\ln(b_1\Lambda)}
+\ln\frac{\ln(t/\Lambda)}{-\ln(b_2\Lambda)}\right],
\label{eq:sgh}
\eeq
where the function $s(q,b)$ are defined in the Appendix A of
Ref.\cite{luy01}.

The hard scale $t_i$'s appeared in Eqs.~(\ref{eq:ab-01}) to (\ref{eq:cd-05})
are of the form
\beq
t_{a} &=& {\rm max}(\sqrt{x_2} m_B,\sqrt{x_1 x_2}m_B,1/b_1,1/b_2)\;,\non
t_{a}^{\prime} &=& {\rm max}(\sqrt{x_1}m_B,\sqrt{x_1 x_2}m_B,1/b_1,1/b_2)\;,\\
t_{b} &=& {\rm max}(\sqrt{x_2|1-x_3-x_1|}m_B,\sqrt{x_1 x_2}m_B,1/b_1,1/b_3)\;,\non
t_{b}^{\prime} &=& {\rm max}(\sqrt{x_2|x_3-x_1|}m_B,\sqrt{x_1 x_2}m_B,1/b_1,1/b_3)\;,\\
t_{c} &=& {\rm max}(\sqrt{(1-x_2) x_3}m_B, \sqrt{|x_1-x_3|(1-x_2)}m_B,1/b_1,1/b_3)\;,\non
t_{c}^{\prime} &=& {\rm max}(\sqrt{|1-x_2(1-x_3-x_1)|}m_B,
    \sqrt{(1-x_2) x_3} m_B,1/b_1,1/b_3)\;,\\
t_{d} &=&{\rm max}(\sqrt{(1-x_2)x_3}m_B,\sqrt{(1-x_2)}m_B,1/b_2,1/b_3)\;, \non
t_{d}^{\prime} &=&{\rm max}(\sqrt{(1-x_2)x_3} m_B,\sqrt{x_3}m_B,1/b_2,1/b_3).
\eeq
\end{appendix}


\end{document}